\documentclass{elsarticle}

\newfont{\toto}{msbm10 at 12 pt}

\newcommand{\bc}{\begin{center}}
\newcommand{\ec}{\end{center}}

\newcommand{\bd}{\begin{description}}
\newcommand{\ed}{\end{description}}

\newcommand{\bi}{\begin{itemize}}
\newcommand{\ei}{\end{itemize}}

\newcommand{\benu}{\begin{enumerate}}
\newcommand{\eenu}{\end{enumerate}}

\newcommand{\bq}{\begin{quote}}
\newcommand{\eq}{\end{quote}}

\newcommand{\be}{\begin{equation}}
\newcommand{\ee}{\end{equation}}
\newcommand{\bea}{\begin{eqnarray}}
\newcommand{\eea}{\end{eqnarray}}

\newcommand{\T}{T}
\newcommand{\dt}{{\Delta t}}

\newcommand{\pii}{{\partial_i}}
\newcommand{\pj}{{\partial_j}}
\usepackage{graphicx}
\usepackage{color}
\usepackage{listings}
\usepackage{bm}
\usepackage{url}
\usepackage{booktabs}
\usepackage{color,soul}
\usepackage[textsize=tiny]{todonotes}

\definecolor{g90}{gray}{.90}
\definecolor{CadetBlue}{rgb}{0.28,0.23,0.54}
\definecolor{OliveGreen}{cmyk}{0.64,0,0.95,0.40}
\definecolor{Brown}{cmyk}{0,0.81,1,0.60}

\usepackage{amssymb}
\begin{document}

\begin{frontmatter}

\title{Massively parallel Lattice Boltzmann codes \\on large GPU clusters}

\author[lab1]{E.~Calore}
\author[lab2]{A.~Gabbana}
\author[lab3]{J.~Kraus}
\author[lab2]{E.~Pellegrini}

%\author[lab1]{S.~F.~Schifano\footnote{Corresponding author. Tel/Fax:+390532974614, email: schifano@fe.infn.it}}

\author[lab1]{S.~F.~Schifano\corref{cor1}}
\cortext[cor1]{Corresponding author. Tel/Fax:+390532974614. Email: schifano@fe.infn.it}

\author[lab1]{R.~Tripiccione}

\address[lab1]{Universit\`a di Ferrara and INFN-Ferrara, via Saragat 1, I-44122 Ferrara, ITALY}
\address[lab2]{Universit\`a di Ferrara, via Saragat 1, I-44122 Ferrara, ITALY}
\address[lab3]{NVIDIA GmbH, Adenauerstr. 20 A4 D-52146 W\"urselen, GERMANY}

%%%%%%%%%%%%%%%%%%%%%%%%%%%%%%%%%%%%%%%%%%%%%%%%%%%%%%%%%%%%%%%%%%%%%%%%
  
\begin{abstract}

This paper describes  a massively
parallel code for a state-of-the art thermal Lattice Boltzmann method. 
Our code has been carefully optimized for performance on one
GPU and to have a good scaling behavior extending to a large number of GPUs.
Versions of this code have been already used for large-scale studies of
convective turbulence.

GPUs are becoming increasingly popular in HPC applications, as they are 
able to deliver higher performance than traditional processors.
Writing efficient programs for large clusters is not an easy task 
as codes must adapt to increasingly parallel architectures, and the 
overheads of node-to-node communications must be properly handled.

We describe the structure of our code, discussing several key
design choices that were guided by theoretical models of performance 
and experimental benchmarks. 
We present an extensive set of performance measurements and identify the 
corresponding main bottlenecks; finally we compare the results of our GPU 
code with those measured on other currently available high performance processors.
Our results are a production-grade code able to deliver a sustained
performance of several tens of Tflops as well as a design and optimization
methodology that can be used for the development of other high performance
applications for computational physics.

\end{abstract}

%%%%%%%%%%%%%%%%%%%%%%%%%%%%%%%%%%%%%%%%%%%%%%%%%%%%%%%%%%%%%%%%%%%%%%%%

\begin{keyword}
     Lattice Boltzmann 
\sep GPU Accelerators 
\sep Massively Parallel Programming 
\sep Heterogeneous systems 
%\sep High Performance Computing 
%\sep Performance Analysis

%% MCS codes here, in the form: \MCS code \sep code
%% or \MCS[2008] code \sep code (2000 is the default)
\end{keyword}

%%%%%%%%%%%%%%%%%%%%%%%%%%%%%%%%%%%%%%%%%%%%%%%%%%%%%%%%%%%%%%%%%%%%%%%%

\end{frontmatter}

%%%%%%%%%%%%%%%%%%%%%%%%%%%%%%%%%%%%%%%%%%%%%%%%%%%%%%%%%%%%%%%%%%%%%%%%

\section{Overview}
High Performance Computing (HPC) has seen in recent years an increasingly  large
role played by Graphics Processing Units (GPUs), offering a  performance level
significantly larger than traditional processors.  GPUs have many slim
processing units on a single chip and perform in  parallel a very large number
(${\cal O}(1000)$) of operations on a correspondingly large number of operands. 
While not limited to such cases, this structure is obviously efficient for 
algorithms offering a large amount of available parallelism; in these cases 
it is possible to identify and concurrently schedule many operations on data 
items that have no dependencies among them.
This is often the case for so-called {\em stencil} codes.  Stencil
codes are typically used to model systems defined on regular lattices;  they
process data elements associated to each lattice site applying some regular
sequence of mathematical operations to data belonging to a fixed pattern  of
neighboring cells. General implementation and optimization of stencils on GPUs 
has been extensively studied by many authors, \cite{stencil1,stencil2,stencil3,stencil4}.
This approach is appropriate also for several computational 
{\em Grand Challenge} applications, such as Lattice QCD (LQCD), or  Computational
Fluid-dynamics using the Lattice Boltzmann method (LBM).  Correspondingly, a
large effort has gone in recent years in porting and optimizing for GPUs codes
and libraries relevant for these applications~\cite{cudagpucode,QUDA,toelke10,bernaschi10}. 

Interesting results have been reported, exhibiting significant 
performance levels obtained on one or just a small number of GPUs. 
However, the number of very large scale computational applications heavily 
relying on GPUs is still limited, partly because the high performance of GPUs 
makes node-to-node communication bandwidth in a large machine a performance 
bottleneck sooner (i.e. fewer nodes) than other platforms, limiting scaling 
behavior on a large number of nodes.

In the last few years, we have conducted a large and systematic analysis 
of several properties of convective turbulence, using as our computational 
tool a massively parallel GPU-based LBM code; physics results have been 
reported elsewhere (see~\cite{ripesi14} and references therein).
After an early development, see \cite{iccs10,ppam11}, and in parallel with 
physics simulations, our code has undergone a systematic process of further 
refinements, improving optimization strategies, adapting to new GPU generations 
and exploiting improved GPU-to-GPU communications tools. 

In this paper we cover the computational aspects of this work, discussing 
the structure of the code, the optimization strategies 
appropriate to boost performances on just one GPU, and the possible approaches 
to improve the scaling behavior of the code on a large GPU cluster; 
we study and analyze in details a number of issues related to state-of-the-art 
computing systems based on GPUs, and identify the corresponding ways-out;
in other words, what we offer here is an attempt at building a sound optimization 
approach for GPUs; while our analysis is based on a specific (but computationally 
relevant) application, we trust that our results may provide useful guidance 
for those adapting and optimizing a wider class of computational applications 
for GPU-based computing.

Analyses on the best options to port LBM codes on massively parallel systems 
have recently appeared~\cite{single-proc}, and detailed studies have focused on the
impact on performance of several memory allocation and access 
strategies~\cite{Pohl,Wittmann,shet1,shet2}. 
Comparisons of results on several multi-core processors and GPUs have also been 
presented in~\cite{sbac-pad13,iccs11,caf11,ppam13}. 
Here we improve and extend those results, further optimizing the codes and 
exploiting recent improvements in GPU-to-GPU data exchange.

This paper is structured as follows: the next section describes the LBM model
that we consider; the following section reviews the architecture of GPU
processors and GPU-based systems. This is followed  by a detailed analysis of
our optimization work, divided in two successive  sections, considering first
the single GPU case, and  then parallelization on a large number of GPUs. We
then discuss  our performance results, including a comparison with
similar codes  optimized for different CPU architectures; our conclusions  and
outlook end the paper. An appendix collects and annotates several critical
code segments, better documenting our implementation choices.

%%%%%%%%%%%%%%%%%%%%%%%%%%%%%%%%%%%%%%%%%%%%%%%%%%%%%%%%%%%%%%%%%%%%%%%%%%

\section{Lattice Boltzmann methods}

In this section, we sketchily introduce the computational method that we adopt,
based on an advanced thermal Lattice Boltzmann scheme. LBM methods (see, e.g.~\cite{sauro} 
for an extended introduction) are discrete in
both position and momentum spaces; they are based on the synthetic dynamics of
{\em populations} sitting at the sites of a discrete lattice. 

This computational method simulates the behavior of a compressible
gas/fluid.
The Thermal-Kinetic description of a compressible gas/fluid of
variable density, $\rho$, local velocity ${\bm u }$, internal energy,
${\cal K}$ and subject to a local body force density, ${\bm g}$, is
given by the following equations:
\begin{eqnarray}
 \partial_t \rho + \partial_i (\rho u_i) = 0 \\ 
 \partial_t (\rho u_k) + \partial_i (P_{ik}) = \rho g_k \\
 \partial_t {\cal K} + \frac{1}{2} \partial_i q_i = \rho g_i u_i
\end{eqnarray}
where $P_{ik}$ and $q_i$ are the momentum and energy fluxes.

In the continuum, one shows that it is possible to recover these
equations, starting from a continuum Boltzmann Equations and introducing a
suitable shift of the velocity and temperature fields entering in the local
equilibrium \cite{JFM},
  $f^{(eq)}({\bm \xi}; \rho, T, {\bm u}) \rightarrow {
   f}^{(eq)}({\bm \xi}; \rho, {\bar T}, {\bm {\bar u}})$.  
The new Boltzmann formulation is then:
 \begin{eqnarray} 
 \label{MASTERSHIFT3} 
 \frac{\partial f }{\partial
     t}+{\bm \xi} \cdot {\bm \nabla} f=-\frac{1}{\tau}(f- {\
     f}^{(eq)})\;\\
     { f}^{(eq)}({\bm \xi}; \rho, \bar T, \bar
   {\bm u})=\frac{\rho}{(2 \pi {\bar T})^{D/2}} e^{-|{\bm \xi}- {\bm
       {\bar u}}|^2/2 {\bar T}},
\end{eqnarray}
and the shifted local velocity and temperature take the
 following form 
 $ \bar {\bm u} = {\bm u} + \tau {\bm g},~~ \bar T
 = \T - \tau^2 g^2/D$ ($D$ is the space dimensionality).

\begin{figure*}[!t]
%\centering
\begin{minipage}{0.45\textwidth}
\includegraphics[width=0.9\textwidth]{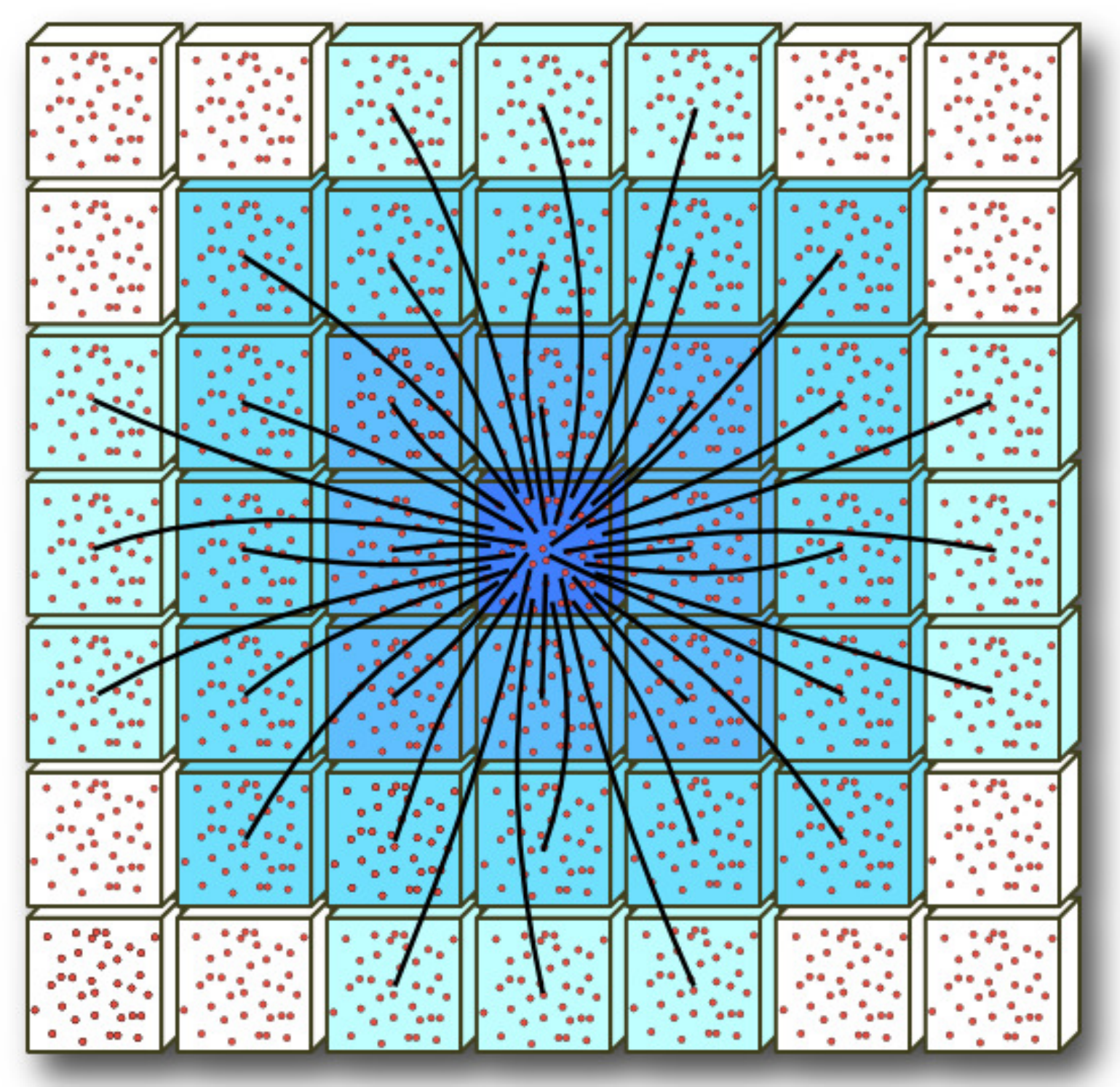}
\end{minipage}
\hspace*{5mm}
\begin{minipage}{0.45\textwidth}
\includegraphics[width=0.9\textwidth]{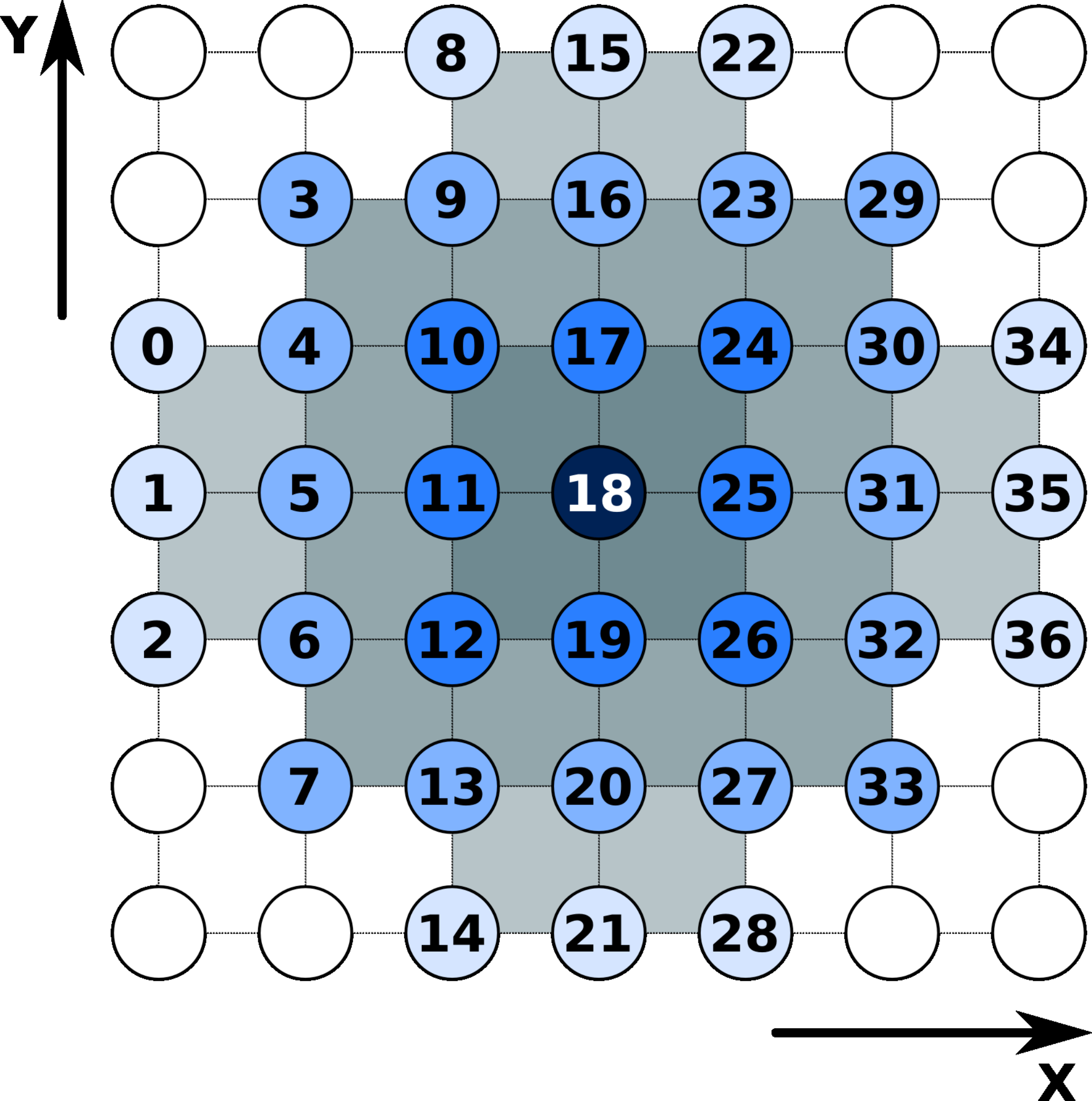}
\end{minipage}
\caption{Left: Velocity vectors for populations in the D2Q37 model,
associated to the lattice hop that they perform in the {\em propagate} phase.
Right: each population is identified by an arbitrary label.}
\label{streamscheme}
\end{figure*}

The discretized counterpart of the continuum description (that we use in
this paper) uses a set of fields $f_l(x,t)$ associated to the so-called {\em
populations}; the latter can be visualized as pseudo-particles moving in
appropriate directions on a discrete mesh (see \figurename~\ref{streamscheme}). 
In this paper we consider a 2D LBM algorithm, that uses 37 populations
(a so called D2Q37 model), recently developed in~\cite{JFM,POF}.
The master evolution equation in the discrete world is:
\begin{equation}
f_{l}({\bm x}+ {\bm c}_{l} \dt,t+\dt) - f_{l}({\bm x},t) =
-\frac{\dt}{\tau}\left(f_{l}({\bm x},t) - f_l^{(eq)}\right);
\label{eq:discrete}
\end{equation}
subscript $l$ runs over the discrete set of velocities, ${\bm c}_l$
(see again \figurename~\ref{streamscheme}) and
equilibrium is expressed in terms of hydrodynamical fields on the
lattice, $ {f}_{l}^{(eq)} = {f}_{l}^{(eq)}({\bm x},\rho,\bar{\bm u},\bar\T)$.
 
To first approximation, the macroscopic fields are defined in terms of the
lattice Boltzmann populations: $\rho = \sum_l f_l$, $\rho {\bm u} = \sum_l {\bm
c}_l f_l$, $D \rho \T = \sum_l \left|{\bm c}_l - {\bm u}\right|^2 f_l$.
When going into all mathematical details, one finds that shifts and
renormalizations have to be applied to the averaged hydrodinamical quantities
to correct for lattice discretization effects. After performing these
manipulations, one recovers  the correct
thermo-hydrodynamical equations:
\begin{eqnarray}
D_t \rho = - \rho \pii u^{(H)}_i \\
\rho D_t u^{(H)}_i = - \pii p - \rho g\delta_{i,3} + \nu \partial_{jj} 
u^{(H)}_i  \\
\rho c_v D_t T^{(H)} + p \pii u^{(H)}_i = k \partial_{ii} T^{(H)} 
\end{eqnarray}
where we have introduced the material derivative, $D_t = \partial_t +
u_j^{(H)} \pj$, neglected viscous dissipation in the heat
equation and the superscript $H$ denotes the lattice-corrected
quantities; $c_v$ is the specific heat at
constant volume  for an ideal gas $p=\rho T^{(H)}$,
$\nu$ and $k$ are the transport coefficients.

%Over the years, several LB models have been developed,
%describing flows in 2 or 3 dimensions, and using sets of populations of
%ifferent size (a model in $x$ dimensions based on $y$ populations is usually
%abeled as $DxQy$).

The LBM model considered in this paper correctly
reproduces the thermo-hydrodynamical equations  of motions of a fluid in two
dimensions, and automatically enforces the  equation of state of a perfect gas
($p = \rho T$).
This is a substantial improvement over simpler LBM schemes in two or three
dimensions (e.g., D2Q9 or D3Q19) that regard the fluid as
incompressible, and introduce ad hoc approximations (e.g., the
Boussinesq approximation) to partially model the dependence of density on
temperature, relevant for convection. 

%%%%%%%%%%%%%%%%%%%%%%%%%%%%%%%%%%%%%%%%%%%%%%%%%%%%%%%%%%%%%%%%%%%%%%%%%%
%
%
%%%%%%%%%%%%%%%%%%%%%%%%%%%%%%%%%%%%%%%%%%%%%%%%%%%%%%%%%%%%%%%%%%%%%%%%%%

An LBM code starts with an initial assignment of the populations, in accordance
with a given initial condition at $t = 0$ on some spatial domain, and then
iterates Eq.~\ref{eq:discrete} for each point in the domain and for as many
time-steps as needed;  at each time step, populations hops from lattice-site to
lattice-site and then incoming populations {\em collide} among one another. In
this step populations mix and their values change accordingly. 
Boundary-conditions are
enforced at the boundary of the integration domain after each time-step
by appropriately modifying the population values at and close to the boundary.\\

From the computational point of view, the LBM approach offers a
huge degree of available  parallelism.
Defining ${\bm y} = {\bm x}+ {\bm c}_{l} \dt$ and rewriting the main
evolution equation as:
\begin{equation}
f_{l}({\bm y}, t+\dt) = f_{l}({\bm y} - {\bm c}_{l} \dt,t) 
-\frac{\dt}{\tau}\left(f_{l}({\bm y} - {\bm c}_{l} \dt,t) - f_l^{(eq)}\right)
\label{eq:master2}
\end{equation}
one easily identifies the overall structure of the computation that evolves the
system by one time step $\dt$; for each point ${\bm y}$ in the discrete grid one:
\benu
\item gathers from neighboring sites the values of the fields $f_l$
      corresponding to populations drifting towards ${\bm y}$ with velocity
      ${\bm c}_l$ and then
\item performs all mathematical processing needed to compute the quantities 
      appearing in the r.h.s. of Eq. (\ref{eq:master2}), for each point in the
      grid.
\eenu
Both steps above are completely uncorrelated 
for different lattice-points, so they can be computed in 
parallel according to any convenient schedule, as long as one makes 
sure that, for all grid points, step 1 is performed before step 2.\\ 
%
%When implementing LBM codes for multi- and many core architectures, 
%the challenge rests in finding the best and most efficient way to 
%match algorithm-level parallelism and available hardware resources.

At each iteration of the loop over time, every lattice-point is processed 
applying in sequence the following three main kernels:
\bi
\item {\tt propagate}: for each lattice-site we move populations according to
       the pattern of~\figurename~\ref{streamscheme} left. This process does not
       perform any mathematics but only  moves blocks of memory
       locations allocated at sparse addresses.  It collects at each site all
       populations that will interact at the next  computational phase
       ({\tt collide}). In this step each site accesses the
       populations of the neighbor cells at distance up to 3 in the 
       grid. 
       %% Wittmann et al. Comparison of different Propagations Steps for LB  
%
\item {\tt bc} adjusts values of the populations at the top and bottom edges of
      the lattice to enforce appropriate boundary conditions (e.g., a constant
      given temperature and zero velocity). This step is necessarily done {\em
      after} propagation, since the latter changes the value of the populations
      close to the boundary points and hence the
      macroscopic quantities that we must keep constant in time.  At the right
      and left boundaries, we apply periodic boundary conditions.  This is most
      easily done by allocating {\em halo columns}, additional  storage where
      copies of the 3 (in our case) rightmost and leftmost columns of the
      lattice are placed before performing the {\tt propagate} step. Points 
      close to the right/left boundaries can then be processed as those in the
      bulk.  If needed, boundary conditions could of course be enforced in the
      same way as we do for the top and bottom edges. 
\item {\tt collide} performs all the mathematical steps associated
      to Eq. \ref{eq:master2} in order to compute the population
      values at each lattice site at the new time step
      (this is called ``collision'', in LBM jargon). Input data for this phase 
      are the populations gathered by the previous {\tt propagate} phase. This 
      step is the truly floating point intensive section of the code; it uses 
      only the population members of the site on which it operates, making the 
      processing of different sites fully uncorrelated.    
\ei

The computational price to
price to pay for this very accurate physics model is that the implementation  of
the steps described above is much more complex than for simpler LBM models. 
More severe computational requirements in terms of memory bandwidth and
floating-point throughput follow. Indeed, {\tt propagate} implies accessing 37 
neighbor cells to gather all populations, while {\tt collide} requires 
$\approx 7000$ double-precision  floating point operations per lattice point.

%%%%%%%%%%%%%%%%%%%%%%%%%%%%%%%%%%%%%%%%%%%%%%%%%%%%%%%%%%%%%%%%%%%%%%%%

\section{NVIDIA GPU Architectures}\label{sec:gpu}

In this work we experiment with two recent generations of NVIDIA GPUs: 
the Tesla processors, C2050 and C2070, based on the GF100 GPU belonging 
to the {\em Fermi} generation, and the latest K20X, K40 and K80 processors, 
based on the {\em Kepler} architecture. The K20X uses a GK110 GPU, 
the K40 a GK110B GPU, and the K80 is a dual GK210 GPU. 
In the following we use interchangeably the name of the systems or that
of the corresponding GPUs, unless ambiguities arise.

NVIDIA GPUs are multi-core processors. 
Processing units are called SM (Streaming Multiprocessors) on Fermi and 
SMX on Kepler (as they have enhanced capabilities). Each processing unit 
has 32 (Fermi) or 192 (Kepler) compute units called CUDA-cores in NVIDIA 
jargon; at each clock-cycle SMs executes multiple warps, i.e. groups of 
32 operations called CUDA-threads which proceed in SIMT fashion~\footnote{
Single Instructions Multiple Threads (SIMT) execution is related 
to SIMD execution but more flexible, e.g. different threads of a 
SIMT group are allowed to take different branches (at a performance penalty).}. 

At variance with CPU threads, context switches among active CUDA-threads 
are instantaneous due to maintaining many thread states. 
Typically one CUDA-thread processes one element of the data-set 
of the application. 
This helps exploit all available parallelism of the algorithm and  
hide latencies by switching among threads waiting for 
data coming from memory and threads ready to run.
This structure has remained stable across both generations. Several enhancements 
are available in the more recent {\em Kepler} processors; for instance, 
{\em Kepler} has 256 32-bit registers addressable by each CUDA-thread
(a $4X$ increase over {\em Fermi}) and each SMX has 65536 registers (a $2X$ increase). 
{\em Kepler} GPUs are also able to increase their clock frequency 
beyond the nominal value, if power and thermal 
constraints allow to do so ({\em GPUBoost}, in NVIDIA jargon). 

Within each generation, minor differences occur: the C2050 and C2070 processors
differ in the amount of available global memory;  the K40 processor has more
global memory than the K20 and slightly improves memory bandwidth and
floating-point throughput; finally the K80 has two enhanced {\em Kepler} 
GPUs with more registers and shared memory than K20/K40 and extended GPUBoost features. 
The Tesla C2050 system has a peak performance of $\approx 1$ Tflops 
in single-precision (SP), and $\approx 500$ Gflops in double-precision (DP); 
on the {\em Kepler} K20 and K40, the peak SP (DP) performance is
$\approx 5$ Tflops ($\approx 1.5$ Tflops), while on the K80 the 
aggregate performance of the two GPUs delivers a peak SP (DP) 
of $\approx 5.6$ Tflops ($\approx 1.9$ Tflops).

Fast access to memory strongly correlates with performance: peak bandwidth 
is $144$ GB/s for the C2050 and C2070 processors, and $250$ and $288$ GB/s 
respectively for the K20X and the K4 0;on the K80, the aggregate peak is 
$480$ GB/s. 
The memory system has an error detection and correction system (ECC) to
increase reliability when running large codes. 
We have always used this feature, even if it slightly reduces available 
memory and bandwidth (e.g. on the Tesla C2050 available memory is reduced 
by $\approx 12\%$; for the propagate kernel measured bandwidth is reduced 
by $\approx 20 \cdots 25$\% ).
For a more complete description, see \cite{fermi,kepler}; 
\tablename~\ref{gpu-evolution} summarizes just a few relevant parameters.
 
%%%%%%%%%%%%%%%%%%%%%%%%%%%%%%%%%%%%%%%%%%%%%%%%%%%%%%%%%%%%%%%%%%%%%%%%
\begin{table}
\caption{Selected hardware features of the GPU systems considered in this paper: 
the C2050 and C2070 are based on the {\em Fermi} architecture, 
while the K20X, K40 and K80 follow the {\em Kepler} architecture.
}
\label{gpu-evolution}
\centering
\resizebox{\textwidth}{!}{
\begin{tabular}{lrrrrrl}
\toprule
                               &  C2050 / C2070  & K20X  & K40      & K80 &  \\
\midrule
GPU                            &   GF100         & GK110 & GK110B   & GK210 &\hspace{-1em} $\times$ 2\\
Number of SMs                  &   16            & 14    & 15       & 13    &\hspace{-1em} $\times$ 2\\
Number of CUDA-cores           &   448           & 2688  & 2880     & 2496  &\hspace{-1em} $\times$ 2\\ 
Nominal clock frequency (MHz)  &   1.15          & 735   & 745      & 562   &                        \\
Nominal DP performance (Gflops)&   515           & 1310  & 1430     & 935   &\hspace{-1em} $\times$ 2\\
Boosted clock frequency (MHz)  &   --            &  --   & 875      & 875   &                        \\
Boosted DP performance(Gflops) &   --            &  --   & 1660     & 1455  &\hspace{-1em} $\times$ 2\\
\midrule     
Total available memory (GB)    &   3 / 6         & 6     & 12       & 12    &\hspace{-1em} $\times$ 2\\
Memory bus width (bit)         &   384           & 384   & 384      & 384   &\hspace{-1em} $\times$ 2\\
Peak mem. BW (ECC-off) (GB/s)  &   144           & 250   & 288      & 240   &\hspace{-1em} $\times$ 2\\
%Peak mem. BW (ECC-on) (GB/s)   &   $\approx$~115 & $\approx$~225    & xxx   & ??                    \\
\midrule
Max Power (Watt)               &   215           & 235   & 235      & 300   &                        \\
\bottomrule
\end{tabular}
}
\end{table}

%%%%%%%%%%%%%%%%%%%%%%%%%%%%%%%%%%%%%%%%%%%%%%%%%%%%%%%%%%%%%%%%%%%%%%%%%
% Aspetti di programmazione delle GPU

We have developed all our codes using CUDA-C~\cite{cuda}, a GPU-specific 
programming language with several features intended to help exploit the parallelism 
available in the algorithm.
A CUDA-C program consists of one or more functions that run either on
the host, a standard CPU, or on a GPU. 
Functions with no (or limited) parallelism run on the host, 
while those exhibiting a large degree of data parallelism go onto the GPU. 
A CUDA-C program is a modified {\tt C} (or {\tt C++}) program including 
keyword extensions defining data parallel functions, called {\em kernels}.
Kernel functions typically translate into a large number of threads, i.e. 
a large number of independent operations processing independent data items. 
Threads are grouped into blocks which in turn form the 
execution {\em grid}. 
The grid can be configured as a 1-, 2- or 3-dimensional array of blocks, 
each block is itself a 1-, 2- or 3-dimensional array of threads, running 
on the same SM, and sharing data through a fast shared memory.
When all threads of a kernel complete their execution, the corresponding 
grid terminates. CUDA-threads run in parallel with 
CPU threads, so it is possible to overlap in time processing on the host and 
the accelerator.
For our purposes this is useful to concurrently schedule computation and 
GPU-to-GPU communication.

%%%%%%%%%%%%%%%%%%%%%%%%%%%%%%%%%%%%%%%%%%%%%%%%%%%%%%%%%%%%%%%%%%%%%%%%

\section{Single-GPU Implementation}

In this section we describe data-structures options, the overall 
organization of the code and optimizing features considering only 
one GPU. The extension to a multi-GPU cluster will be considered in the 
next section. 

%%%%%%%%%%%%%%%%%%%%%%%%%%%%%%%%%%%%%%%%%%%%%%%%%%%%%%%%%%%%%%%%%%%%%%%%%%
%
\begin{figure}[t]
\centering
\includegraphics[width=\textwidth]{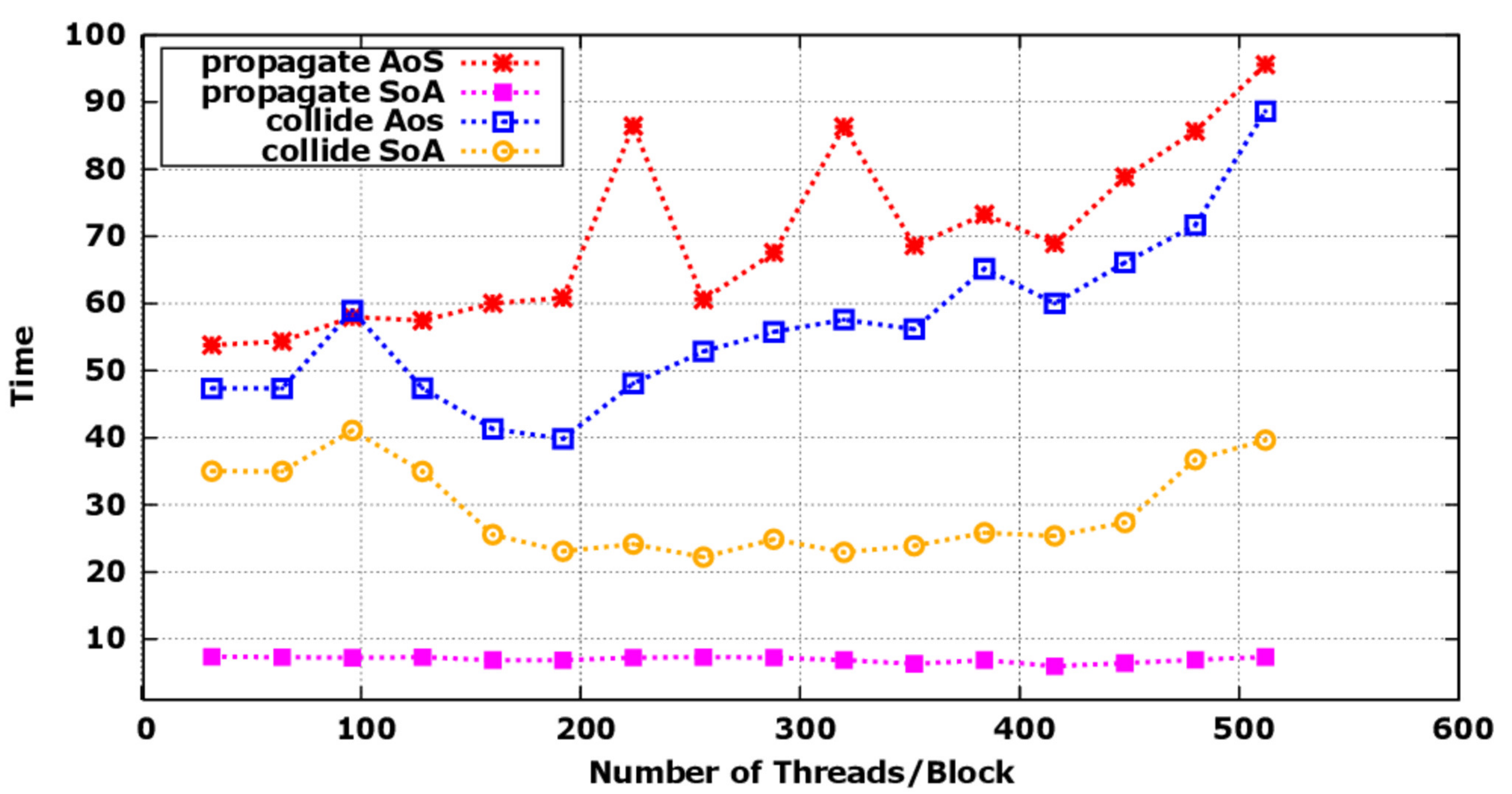}
\caption{Execution times (arbitrary units) of the {\em propagate} 
and {\em collide} kernels as a function of the number of threads per 
block, using the AoS and SoA data layouts.}
\label{fig:aos-vs-soa}
\end{figure}
%
%%%%%%%%%%%%%%%%%%%%%%%%%%%%%%%%%%%%%%%%%%%%%%%%%%%%%%%%%%%%%%%%%%%%%%%%%%

\subsubsection*{Data Structure Analysis}

A major decision affecting the overall structure of the code has to 
do with the choice of an appropriate data organization, which has a strong 
impact on the ability of the system to fetch from memory all the data 
elements needed by the processor. In LBM popular data organizations are 
array of structures (AoS) or structure of arrays (SoA).
With AoS, all populations associated to each 
lattice site are stored one after the other in memory; conversely, SoA 
stores data items corresponding to each population at all lattice sites
one after the other. 
For serial computations on cache-based architectures, like traditional CPUs, 
the AoS scheme is preferable as it improves the locality of populations 
associated to each lattice point, and better suits the cache structure 
and hierarchy of these processors. 
On the other hand SoA is required for data parallelism computation 
typical of GPUs, since it allows to process data associated to several 
lattice sites in parallel and allows coalescing of memory accesses, that 
helps achieve high sustained memory bandwidth.
This is substantiated in figure~\ref{fig:aos-vs-soa}, showing a preliminary
performance analysis of two critical kernels;
{\em propagate} (a memory-bound kernel, see later for details) 
is at least a factor $5X$ faster using the SoA scheme, while {\em collide} 
(a compute-bound kernel) is roughly $2X$ faster.

Having settled for an SoA structure, we store the lattice in column-major 
order in the $Y$ direction (we arbitrarily select one of the two possible choices), 
and keep in memory two copies of the lattice. 
Code sections alternatively read one copy and write on the other copy; 
this technique is known as double-buffering.
This helps maximize parallelism, allowing to map one thread per lattice site, 
and then processing all sites in parallel.

%%%%%%%%%%%%%%%%%%%%%%%%%%%%%%%%%%%%%%%%%%%%%%%%%%%%%%%%%%%%%%%%%%%%%%%%%%
%
\begin{figure}[t]
\centering
\includegraphics[width=0.75\textwidth]{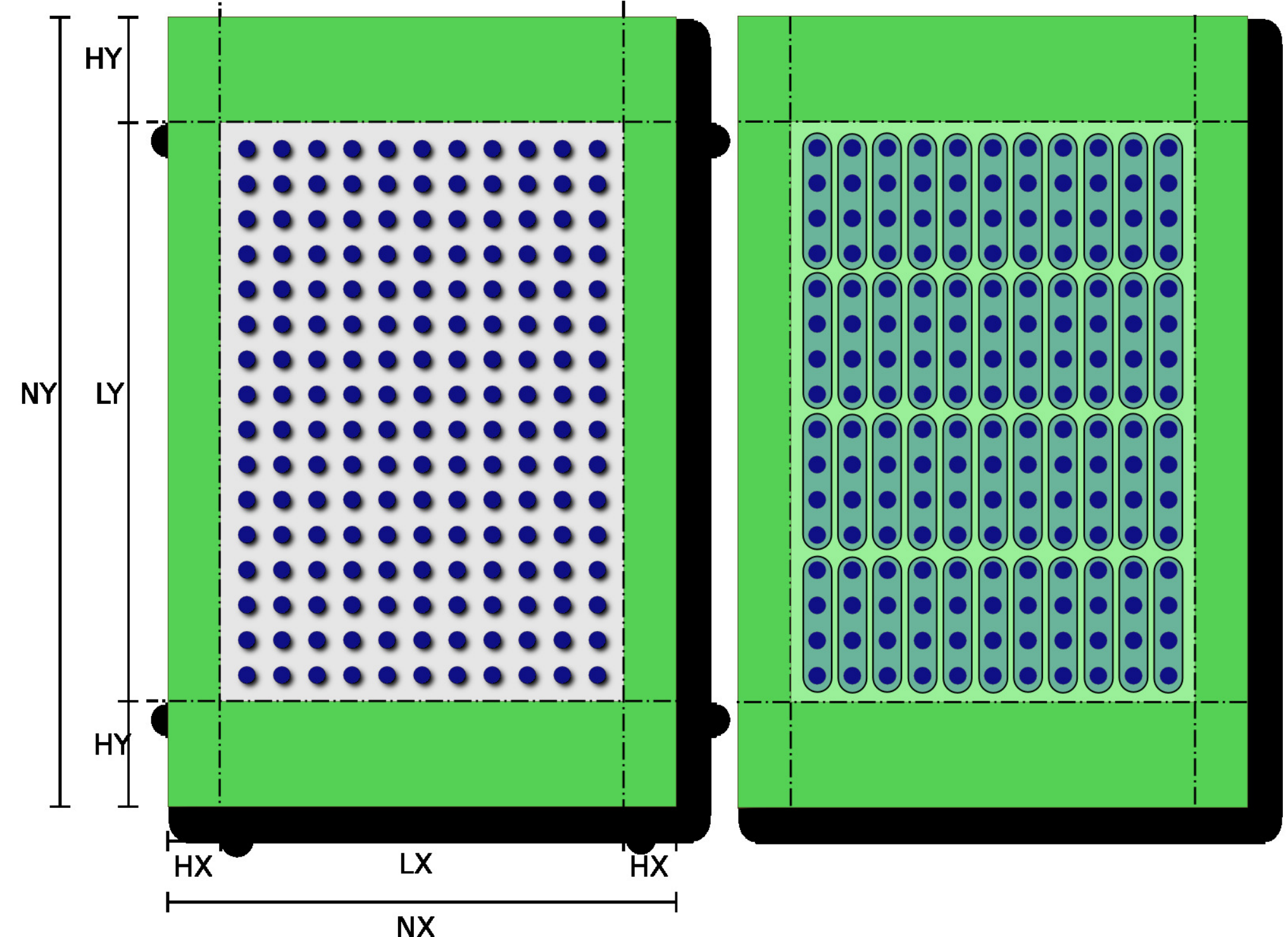}
\caption{Left: Allocation of lattice data in global memory; green regions
are the halo frames, the white area is the physical 
lattice. Right: sketchy view of the mapping of the lattice on 
CUDA thread-blocks and block-grids.}
\label{fig:cudagrids}
\end{figure}
%
%%%%%%%%%%%%%%%%%%%%%%%%%%%%%%%%%%%%%%%%%%%%%%%%%%%%%%%%%%%%%%%%%%%%%%%%%%

We surround the physical lattice by halo-columns 
and rows, see \figurename~\ref{fig:cudagrids}: for
a physical lattice of size $L_x \times L_y$, we allocate 
an array of $NX \times NY$ lattice points, $NX=H_x+L_x+H_x$, and 
$NY=H_y+L_y+H_y$. 
This makes the computation uniform for all sites and avoid 
thread divergences which break data-parallelism and degrade performances. 
The algorithm requires a halo-thickness of just 3 points, since populations 
move up to three sites at each time step. 
It is convenient to use a larger halo thickness in the $Y$ directions 
(Hy = 16), in order to keep data aligned (multiples of 32, the size of 
the warp, permit more efficient access through ``coalescing'') and to 
maintain also cache-line alignment in multiples of 128 Bytes.
 
\subsubsection*{Code Organization}

Our code starts on the host, and at each iteration four main steps 
execute: first the {\tt pbc} kernel update halos, and then three kernels 
-- {\tt propagate}, {\tt bc} and {\tt collide} -- perform the 
required computational tasks. Each kernel corresponds to a CUDA-C function.

%% pbc
For a single GPU implementation the {\tt pbc} kernel updates only 
the left and right halos, as we enforce periodic boundary conditions along $X$; 
this amounts to copying data from the three right-most columns of the physical 
lattice to the left halo columns, and vice-versa. In this case, we 
move data stored at contiguous elements in memory, so we handle it by 
efficient CUDA-C memory-copy library functions. We make two calls to the 
{\tt cudaMemcpyAsync} function; execution overlaps in time, substantially 
increasing performance.

%%%%%%%%%%%%%%%%%%%%%%%%%%%%%%%%%%%%%%%%%%%%%%%%%%%%%%%%%%%%%%%%%%%%%%%%%%
%
\begin{figure}[!t]
\centering
\includegraphics[width=\textwidth]{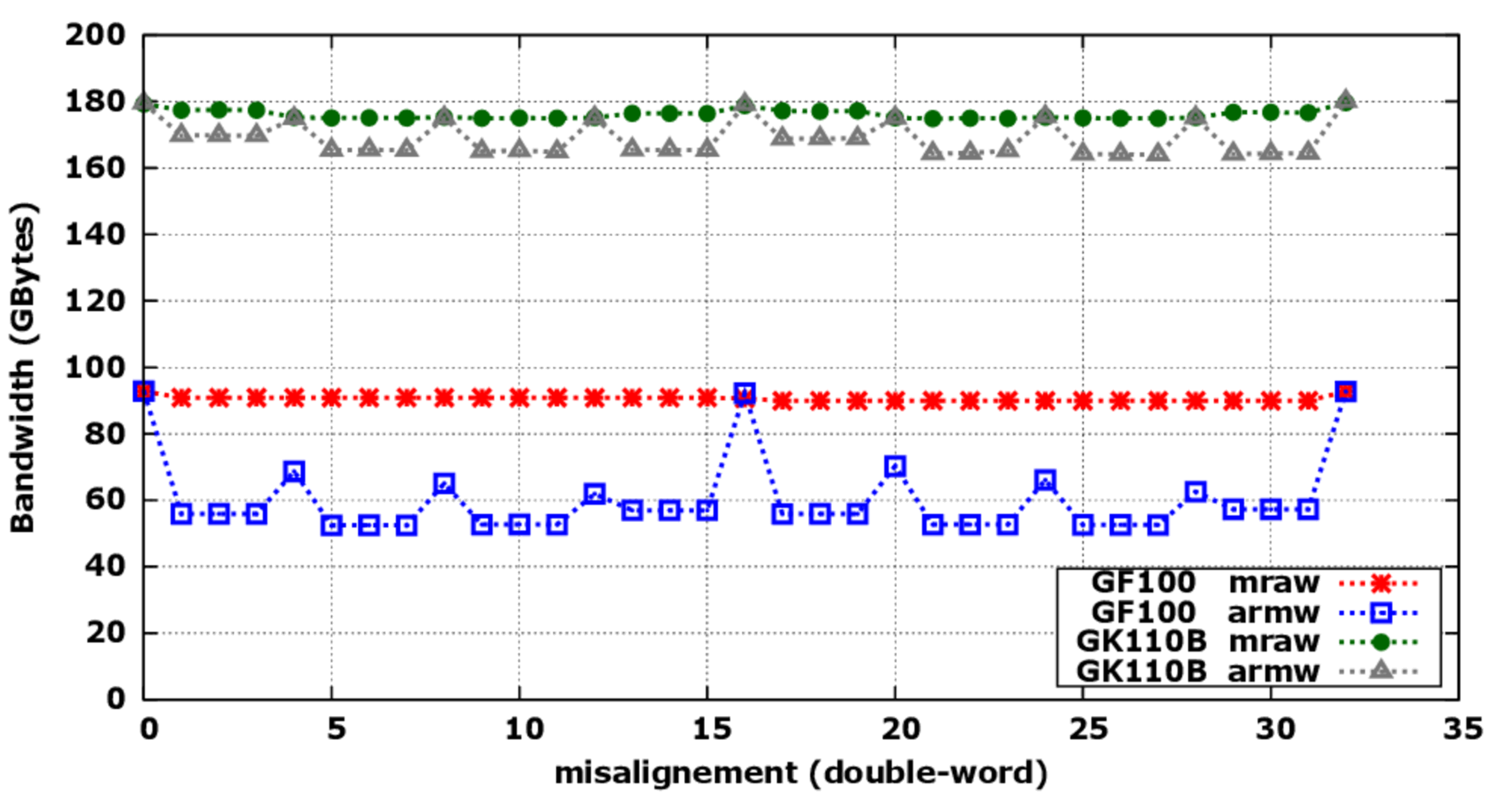}
\caption{
\label{misalignement}
Performance (ECC enabled) of misaligned-read-aligned-write 
(mraw) and aligned-write-misaligned-read (armw) accesses on the C2050 
(GF100 GPU), and K40 (GK110B GPU) systems.
}
\end{figure}
%
%%%%%%%%%%%%%%%%%%%%%%%%%%%%%%%%%%%%%%%%%%%%%%%%%%%%%%%%%%%%%%%%%%%%%%%%%%

\subsubsection*{The {\tt propagate} and {\tt bc} Kernels}
 
%%% propagate
The {\tt propagate} kernel moves populations at each site according to 
the pattern shown in \figurename~\ref{streamscheme}. Two options are possible:
i) {\em push} moves all populations of each lattice site to the appropriate 
neighbor sites; 
or ii) {\em pull} gathers populations from neighbor sites to each destination site. 
{\em push} performs aligned reads and misaligned writes, while the opposite happens in {\em pull}. In 
both cases, misaligned memory operations are needed. 
\figurename~\ref{misalignement} plots the measured bandwidth 
of a memory-copy kernel using misaligned reads  
and aligned writes (mraw) and vice-versa (armw). 
The {\tt mraw} scheme is faster on both GPU generations even if the 
performance gain, large for the C2050 system, is smaller for the 
more recent K20 and K40 GPUs.
This test obviously suggests to adopt the {\em pull} scheme.

\begin{figure}
%\centering
\includegraphics[width=\textwidth]{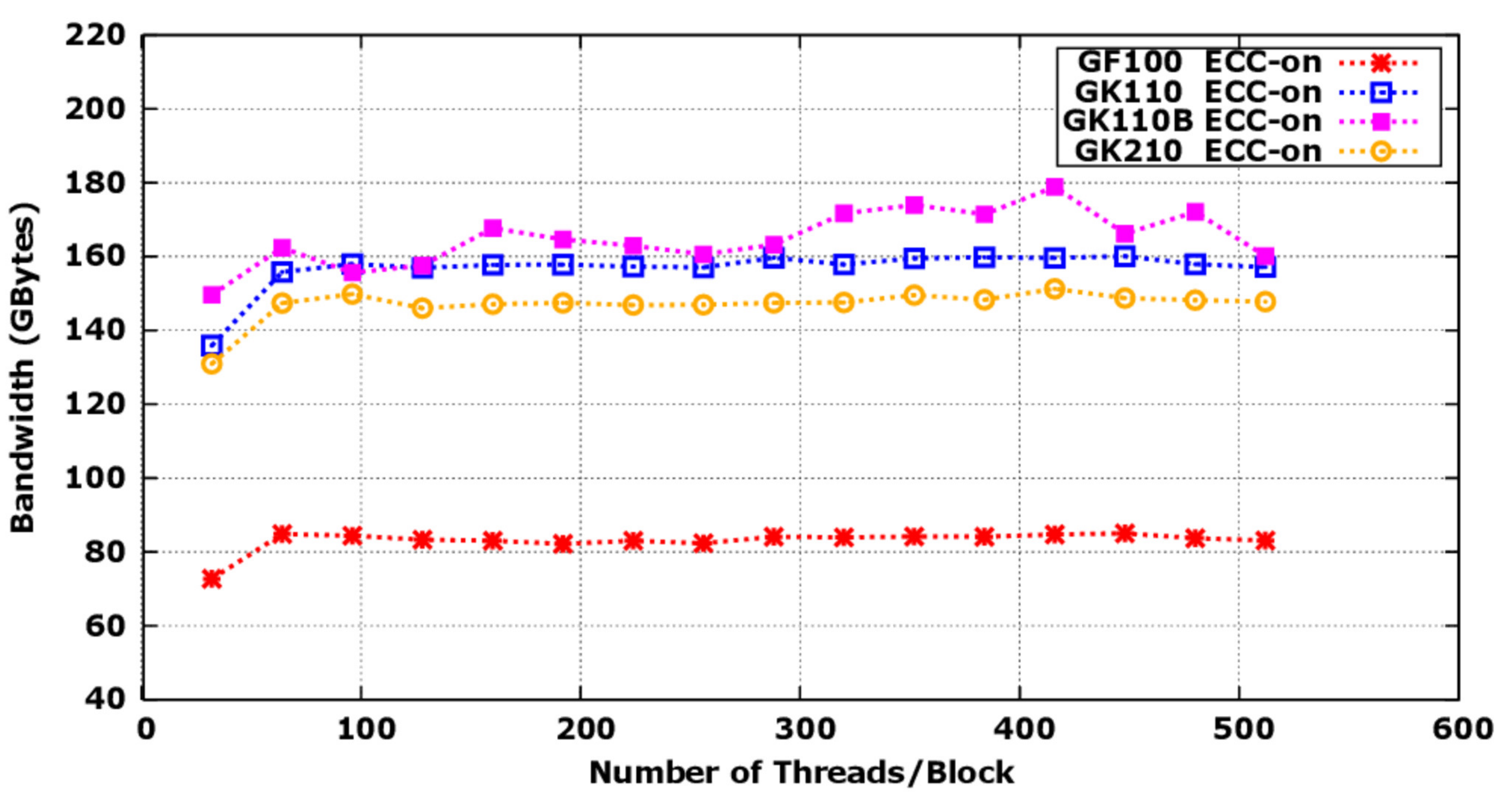}
\caption{
\label{propagate-bench}
Performance of the {\tt propagate} kernel (ECC enabled), 
on C2050 (GF100 GPU), K20X (GK110 GPU), K40 (GK110B GPU) systems 
and on one GPU of a K80 (GK210 GPU) board, vs. number of threads per block, 
on a lattice of $\approx 4$ million of cells.
}
\end{figure}

For this kernel, each CUDA-block is configured as a 
unidimensional array of {\tt N\_THREAD} threads, processing data allocated 
at successive locations in memory, while the grid of blocks is a bi-dimensional 
array of {\tt ($L_y$/N\_THREAD $\times$ $L_x$)} blocks, 
see \figurename~\ref{fig:cudagrids}, right.
{\tt N\_THREAD}, in principle a free parameter, must be accurately 
tuned for performance: {\tt N\_THREAD} should be large enough because 
it translates in long and efficient memory access sequences. On the 
other hand, {\tt ($L_y$/N\_THREAD $\times$ $L_x$)} should also be large, 
because it translates into many independent sequences, so some sequence 
is almost always ready to execute while other are waiting for data 
incoming from memory.
\figurename~\ref{propagate-bench} shows the impact of this parameter 
on performance, displaying the effective memory bandwidth as a function 
of the number of threads per block.
We see that performance
stabilizes to a reasonable level as long as ${\tt N\_THREAD} \ge 64$;
on the C2050 processor, we reach a bandwidth of $\approx 85$ GB/s
that increases to $\approx 180$ GB/s on the K40; on K20X, it is around 
$160$ GB/s and on one GPU of a K80 board is $\approx 150$ GB/s. These figures 
nicely agree with benchmark results presented in figure~\ref{misalignement}.
%For this kernel, if we disable memory ECC support, bandwidth increases 
%approximately by a factor $\approx 1.20$X on all systems; however in our 
%benchmarks and runs we always keep ECC enabled to make simulations reliable 
%against memory failures.

%%%%%%%%%%%%%%%%%%%%%%%%%%%%%%%%%%%%%%%%%%%%%%%%%%%%%%%%%%%%%%%%%%%%%%%%%%
%
%
%%%%%%%%%%%%%%%%%%%%%%%%%%%%%%%%%%%%%%%%%%%%%%%%%%%%%%%%%%%%%%%%%%%%%%%%%%

%%% bc
The {\tt bc} kernel enforces boundary conditions (constant temperature 
and zero velocity of the fluid) at the top and bottom of 
the lattice; it runs only on the 
threads corresponding to lattice sites with coordinate {$y=0, 1, 2$} and 
{$y=L_y-1, L_y-2, L_y-3$}.
The layout of each CUDA block is the same as for the propagate 
kernel, and the code uses {\tt if} statements to disable threads not involved in the 
computation. This causes thread divergence, but, as we show later, 
the computational cost of the {\tt bc} kernel is negligible compared to all 
other steps, so performance drops in this kernel have a minor global impact.

\subsubsection*{The {\tt collide} Kernel}

%%% collide
The {\tt collide} kernel takes care of the collision of populations gathered 
by the {\tt propagate} step. At each time step, each thread reads populations of each lattice site from the {\tt prv} arrays, performs all needed mathematical 
operations and stores the resulting populations onto the {\tt nxt} array.
The roles of {\tt nxt} and {\tt prv} are swapped at each iteration.
In this scheme,  memory reads and writes are always sequential 
and properly aligned, enabling memory coalescing.

{\tt collide} is a strongly compute bound routine. This is shown in Tab. 
~\ref{nvprof:tab}, collecting the output of the NVIDIA {\tt nvprof} 
execution profiler. 
After compilation and optimization the {\tt collide} kernel executes 
$6472$ double-precision mathematical operations for each lattice site, 
but only $\approx 72\%$ are executed as more efficient Fused-Multiply-Add (FMA), 
slightly reducing the overall performance. 
Data from the table translate into an arithmetic intensity of $\approx 11$ 
Flops/byte; using this figure and the ALU utilization, the needed memory 
bandwidth is only one third of the peak available on the GPU. This confirms 
that the kernel is limited by arithmetic throughput, rather than memory bandwidth. 
%
%Data from the table translate into an arithmetic intensity of $\approx 11$ 
%Flops/byte, suggesting that this kernel is limited by arithmetic throughput,
%rather than memory access. 
%
The thread and block organization here is the same as for {\tt propagate}, 
see again \figurename~\ref{fig:cudagrids} right, 
and the corresponding parameters should again be tuned for performance. 
There is tension between the gains arising from a large number of data 
points being processed together and the limited register space available 
to store the huge number of constants and intermediate results that need 
to be maintained inside the processor as different data blocks are processed 
in turn.
%
%%%%%%%%%%%%%%%%%%%%%%%%%%%%%%%%%%%%%%%%%%%%%%%%%%%%%%%%%%%%%%%%%%%%%%%%%%
\begin{table}
%\bc
\centering
\caption{\label{nvprof:tab} Output of the NVIDIA {\tt nvprof} profiler
for {\tt collide} on a lattice of $2048 \times 1024$ sites. 
The table shows the number of ADD, MUL and FMA (fused Multiply-Add) 
operations, the total number of floating-point operations per site 
and the fraction of cycles in which the ALU is used (ALU Utilization).}
\begin{tabular}{lr}
\toprule
FLOPS (Double Add) &  704684032  \\
FLOPS (Double Mul) & 1530290176  \\
FLOPS (Double FMA) & 5629132825  \\
\midrule
FLOPS per site     &      6472   \\
\midrule
ALU Utilization    & High (70\%) \\
\bottomrule
\end{tabular}
%
%\ec
\end{table}
%%%%%%%%%%%%%%%%%%%%%%%%%%%%%%%%%%%%%%%%%%%%%%%%%%%%%%%%%%%%%%%%%%%%%%%%%%%

%
Going into more technical details, we use data prefetch to hide memory 
accesses and on {\em Kepler} all loops accessing the thread-private prefetch 
array have been unrolled via {\tt \#pragma unroll}. 
This allows the compiler to keep the elements 
of the prefetch array in registers exploiting the larger register 
file available on the {\em Kepler} GPUs. 
We experimentally searched for the best tradeoff between register spilling 
and device occupancy manually setting the maximum number of threads per block
and the minimum number of blocks per SM. This can be done using the 
{\tt launch\_bounds } directive~\cite{cuda}; this handcrafted optimization
step  improves performances by $\approx 20\%$.
 
%%%%%%%%%%%%%%%%%%%%%%%%%%%%%%%%%%%%%%%%%%%%%%%%%%%%%%%%%%%%%%%%%%%%%%%%%%
%
\begin{figure}[!t]
%\centering
\includegraphics[width=\textwidth]{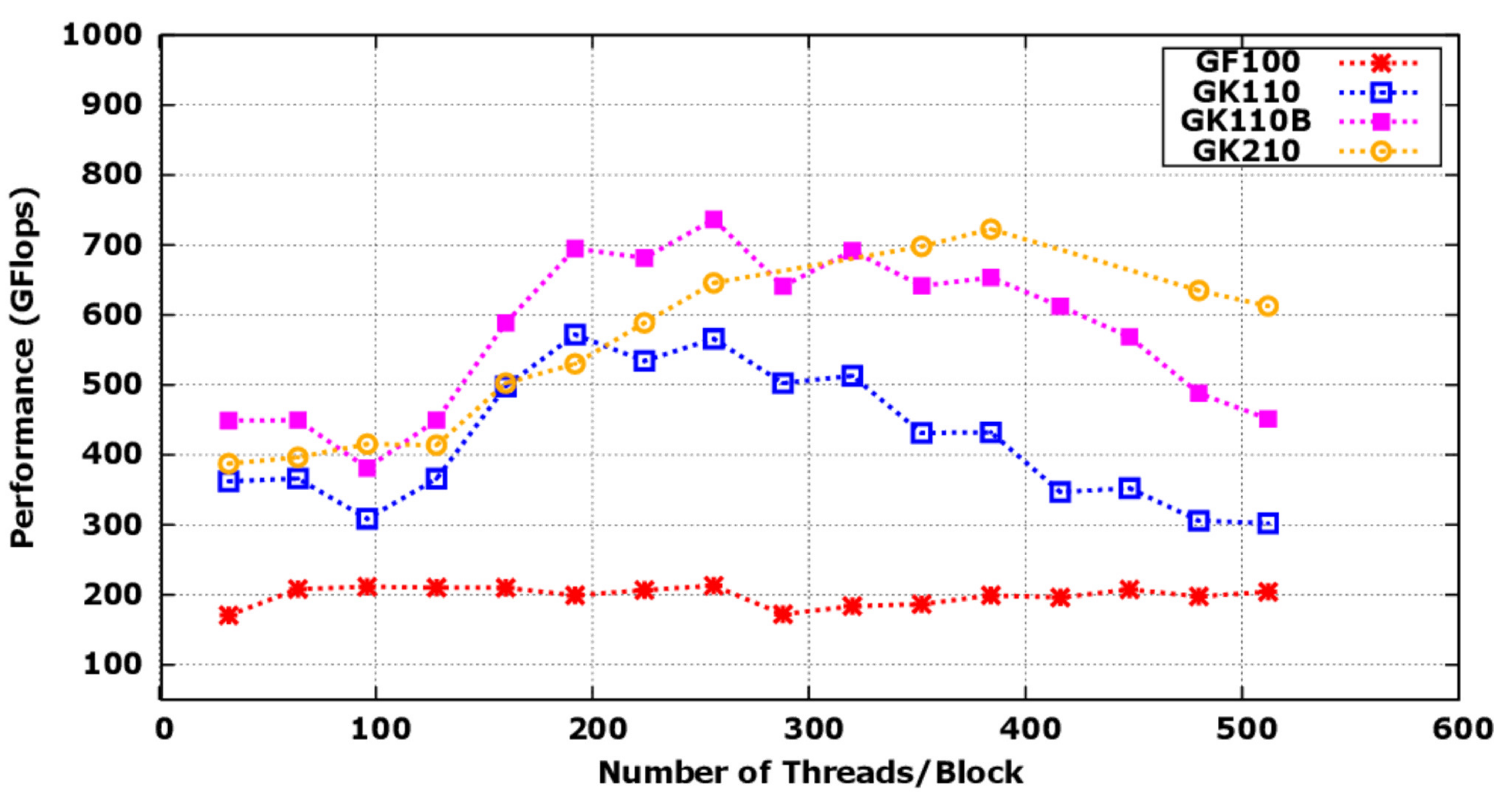}
\caption{Performance of the collide kernel on C2050 (GF100), K20X (GK110), 
K40 (GK110B) systems and on one GPU of a K80 (GK210) board vs. 
the number of threads per block, on a lattice of $\approx 4$ million of 
cells.}
\label{collide-bench}
\end{figure}
%
%%%%%%%%%%%%%%%%%%%%%%%%%%%%%%%%%%%%%%%%%%%%%%%%%%%%%%%%%%%%%%%%%%%%%%%%%%

\subsubsection*{Performance Analysis}

\figurename~\ref{collide-bench} shows the performance of the {\tt collide} kernel 
as a function of the number of threads per block. 
On the C2050, it basically reaches a plateau for a number of threads 
larger than 64, and the sustained performance is 
$\approx 210$ GF/s, that is $\approx 40\%$ of peak. 
On the K20X and K40, the behavior is different: 
performance improves up to 256 threads per block reaching a peak value of 
$\approx 570$ GF/s for the K20X and $\approx 730$ GF/s for the K40, 
that is, respectively $\approx 43\%$ and $\approx 51\%$ of 
peak. 
Finally on one GPU of the K80 board, top performance is the same as the K40, but it is 
obtained with a larger value of threads per block, and performance 
decreases less sharply if this parameter is further increased.

As we try to use a larger number of threads, performance drops because the 
number of needed registers is larger than the available resources on the SMs.
%
%Table~\ref{nvprof:tab} however shows that the usage of arithmetic pipeline is
%$\approx 70\%$. As the aggregate (read+write) memory bandwidth  measured by the
%profiler is around $69$~GB/s, and the load store sub-system utilization is
%reported as "Low" -- around 10\% -- this cannot be explained by the %capabilities 
%of the memory system. 
%
%The performance drop with larger number of threads on Kepler can be explained 
%by limited resources on the SM. 
%
Indeed, as already remarked, the {\em Kepler} version of the {\tt collide} 
kernel holds the values of the prefetch array in registers. Since the size of 
the register file is limited, more and more 
registers must be spilled to global memory if more threads per block are used. 
The L1 cache is too small to handle all spills and although the available device 
memory bandwidth for the spilling is not a bottleneck this has a negative impact 
on performance. 
%
%%We therefore conclude that additional instructions necessary 
%%to handle the registers spills and the longer latencies they incur are a 
%%performance limiter for the {\tt collide} kernel. 
%
%
The larger register file of the enhanced Kepler SMX on the {\tt Tesla} K80 helps with that: 
this is why on this processor performance is more stable as the number 
of threads per block increases.
%
%In conclusion the collide kernel is limited by memory latencies in handling 
%register spilling and corresponding instruction cache misses.
%
In conclusion, the collide kernel is limited by memory latencies as the 
large amount of state per thread does not allow to run enough threads 
concurrently on the SMs to cover all latencies.

Further optimization steps are possible and have already been discussed in the 
literature, such as fusing the {\tt propagate} and {\tt collide} steps
\cite{Wittmann}; 
also, {\tt pbc} can be overlapped with the execution of these steps, with some 
change in the scheduling of operations. The details of these optimizations 
depend significantly on how the lattice is split across processors in a 
multi-GPU implementations, so we defer this discussion to the next section.
   
%%%%%%%%%%%%%%%%%%%%%%%%%%%%%%%%%%%%%%%%%%%%%%%%%%%%%%%%%%%%%%%%%%%%%%%%
\section{Multi-GPU implementation}

In this section we describe the structure and implementation of our code
for a (large) multi-GPU cluster.
%We consider a 
%general purpose commodity cluster where nodes have two or more GPU installed 
%and are interconnected trough high speed network such as Infiniband. 
%
%This means that communications among GPUs needs to be optimized according 
%to their physical allocation and may require different coding solutions.
%
We divide this section in three parts:
we first discuss some simple theoretical model of performance that have 
guided our parallelization strategies; we then review the programming 
environment and tools available to support GPU-to-GPU communication, 
and finally present details of our implementation.
%the {\em introduction} analyzes the 
%possible ways to split the data-lattice over a grid of GPUs; the 
%second part discusses code solutions that may be used to make communications 
%efficient, and the last part presents our implementation. 

\subsection*{Modelling the impact of communications}

A parallel multi-processor LBM code is in principle 
straightforward: one just maps regular tiles of the physical lattice on 
the processors; the processing load is balanced among processing elements 
if all domains have the same size; finally tile-to-tile 
communication patterns are regular and predictable and only involve 
(logically) nearest-neighbor processors.
Still, node-to-node communications are an unavoidable overhead that may 
become serious, hampering performance scaling of the program, as the number 
of nodes increases. 
The amount of data to be moved is roughly proportional to the surface of each
computational domain, while computing scales as the domain volume, so, in 
order to ensure better scaling figures, one should
i) identify the domain decomposition that minimizes the surface-over-volume ratio
and ii) overlap communications with parts of the computation that have no
dependency with data incoming from neighbor nodes.

Simple performance models may guide actual 
program development. For a lattice with $N$ points in $D$ dimensions 
(i.e., with linear size $L = N^{1/D}$) one maps regular tiles onto $N_p$ processors;
each tile contains points associated to all coordinate values in
$D - d$ dimensions and an equal number of coordinate values in the remaining $d$ 
dimensions ($d \le D$). 
One easily finds that the surface-over-volume ratio ($S/V$) is
\begin{equation}
S/V \simeq d ~ N_p^{1/d},
\end{equation}
so in principle $d  = D$ ($d = 2$ in our case) should have the best scaling
performance. 

In practice, things are more complex for several reasons. 
One relevant point is that communications of data elements corresponding to 
borders in different directions may have widely different bandwidths. 
This depends on the data layout in memory, as this dictates which surface elements 
are stored at non-contiguous addresses, usually at fixed distance 
({\em stride}) from each other. 
For memory-contiguous data words, a node-to-node communication involves a stream of
data items from memory elements to the network interface, and then from the
network interface again to contiguous memory cells. 
However data from sparse memory location has to be first gathered into a contiguous 
buffer, then transmitted and finally scattered to memory cells at sparse addresses. 
These access patterns may be much slower than for contiguous memory cells, so 
effective bandwidths may be widely different. 

Consider a $2D$ lattice of $L_x \times L_y$ sites, that we partition on 
$N_p$ processing elements. 
Each processing element handles a tile of $(L_x/n_x) \times (L_y/n_y)$ sites 
($n_x \times n_y = N_p$). 
Assume that transfers in the $X$ and $Y$ directions have effective (and in general different) 
bandwidths $B_x$ and $B_y$.
The time needed to move information across all boundaries of each domain
is proportional to $T_ {C}$ (through a factor $S$ that counts how many 
bytes have to be moved for each boundary site):
\begin{equation}
T_ {C} = \frac{L_y}{B_y n_y} + \frac{L_x}{B_x n_x},
\label{eq:t2d}
\end{equation}
We now ask what is the optimal choice for $n_x$ and $n_y$ corresponding 
to the minimum of Eq.~\ref{eq:t2d}, with  $n_x \times n_y = N_p$. 
One easily finds that:
\begin{equation}
n_x  =  \sqrt{N_p} R, ~~~~~~~ n_y = \sqrt{N_p} / R
\label{eq:bestn}
\end{equation} 
with $R = \sqrt{\frac{L_x B_y}{L_y B_x}}$ a factor taking into 
account the aspect-ratio of the lattice
and the mismatch of the bandwidth values. Using these optimal choices, 
we further obtain: 
\begin{equation}
T_{C}^{min} = \frac{2}{\sqrt{N_p}} \sqrt{ \frac{L_x L_y}{B_x B_y} }.
\label{eq:t2dmin}
\end{equation}
Total processing time $T$ is the sum of communication time and (on-node) processing 
time $T_P$; the latter grows as the number of lattice sites handled by each processor, 
$T_P = \beta  N/N_p$, so, $T = T_P + T_C$,
\begin{equation}
T = \beta \frac{N}{N_p} \{1 + \frac{4S}{\beta} \frac{1}{\sqrt{B_x B_y}}\sqrt{N_p/N}\}
\label{eq:tot2d}
\end{equation}
In this and in the following equations, we always write $T$ as a scaling term 
($\beta N/N_P$) multiplied by a scale violating one 
\footnote{we use throughout the term ``scaling'' to mean ``linear-scaling behaviour'', 
and the term ``scale violation'' for ``violation of linear scaling behaviour''.} (in braces). 
For comparison, if we tile the lattice in just one dimension (e.g., $Y$), 
a similar reasoning tells us that
\begin{equation}
T = \beta \frac{N}{N_p} \{1 + \frac{2SL_y}{\beta B_y} N_p/N\}
\label{eq:tot1d}
\end{equation}
which has obviously more severe asymptotic (large $N_p$) scaling violations. 

%% THIS IS THE PART RELATED TO THE BRENT THEOREM
It may be interesting to look at Eqs.~\ref{eq:tot2d} and \ref{eq:tot1d} from the 
point of view of Brent's theorem~\cite{brent}, that, in the framework of a PRAM model, 
states that
\begin{equation}
T \le S_N + \frac{W_N - S_N}{N_p},  
\label{eq:brent}
\end{equation}
with $W_N$ the overall number of operations to be performed, and $S_N$ the longest 
path in the dependency graph of the algorithm. In our case, $W_N$ depends 
linearly on the lattice size, $W_N = w\times N$, with $w$ counting the number 
of operations to be performed on each lattice site, while $S_N$ is $N$ independent, 
$S_N = w$, since operations on different points of the lattice have no dependencies 
among them. In this case, Eq. \ref{eq:brent} reads
\begin{equation}
T = w + \frac{w(N -1)}{N_p} = w(1+\frac{N-1}{N_p}) \approx w N/N_p;
\end{equation}
our model (Eqs. \ref{eq:tot2d} and \ref{eq:tot1d}) obtains the same result, 
apart from a correction due to communication overheads, that is not considered 
by Brent's theorem since the PRAM model assumes shared-memory with uniform 
access time.
 
%%%%%%%%%%%%%%%%%%%%%%%%%%%%%%%%%%%%%%%%%%%%%%%%%%%%%%%%%%%%%%%%%%%%%%%%%%%%%
\begin{figure}
\centering
\includegraphics[width=\textwidth]{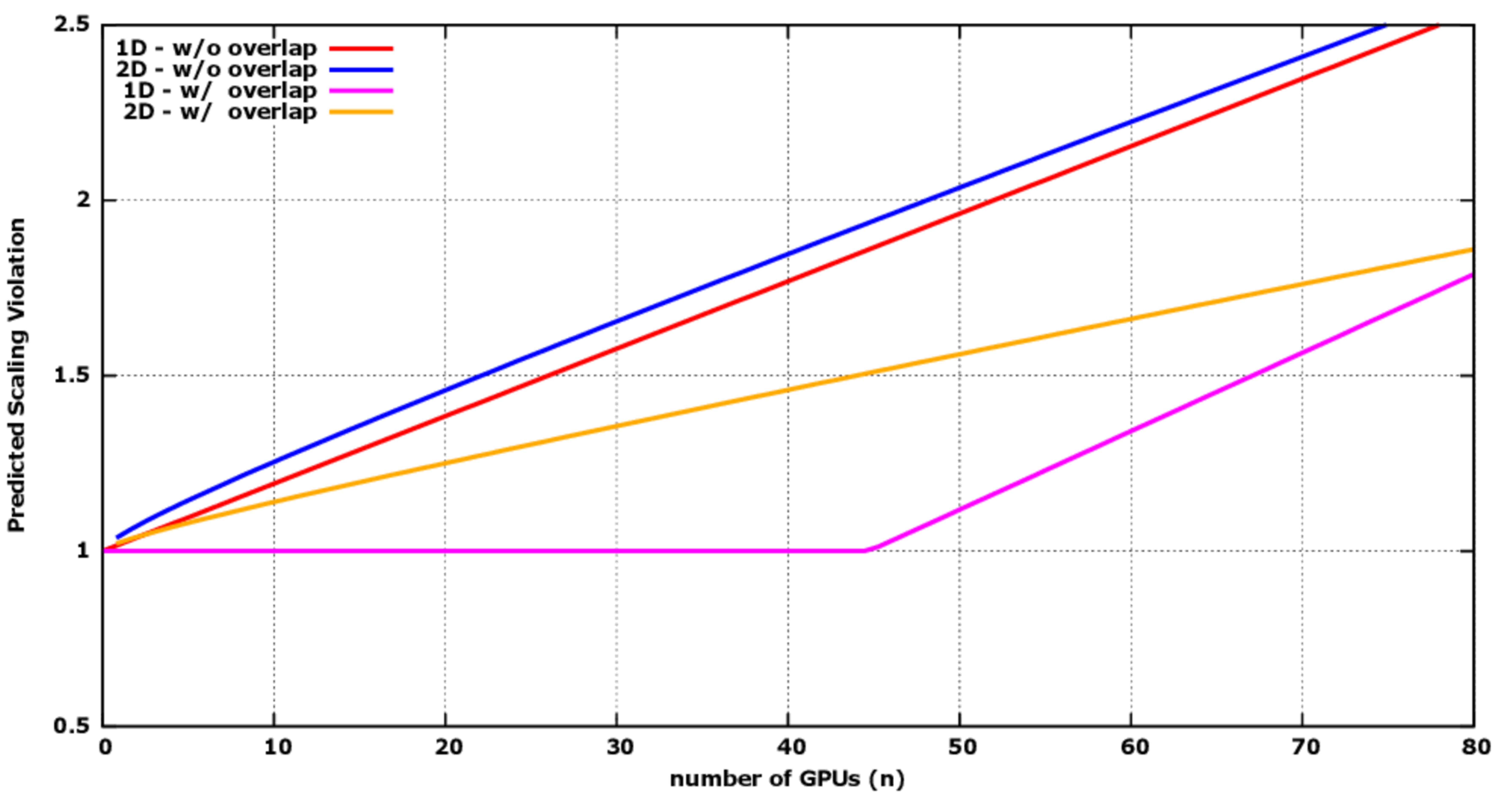}
\caption{Scaling violations predicted by our performance model
(Eqs. \ref{eq:tot2d} to \ref{eq:tot2dov}) on a lattice of $1940 \times 1940$ 
sites, using experimentally measured single-node processing performance and
node-to-node bandwidths for contiguous and non-contiguous data buffers. 
In 2D we use the possibly not-optimal choice $n_x = n_y$. }
\label{plot:model}
\end{figure}
%%%%%%%%%%%%%%%%%%%%%%%%%%%%%%%%%%%%%%%%%%%%%%%%%%%%%%%%%%%%%%%%%%%%%%%%%%%%%

A further key observation is that, for each tile, all lattice points belonging 
to the bulk (i.e., away from the tile boundary by more than $3$ lattice points) 
have no dependency from data of other nodes. This suggests to overlap bulk 
processing and data transfer. For an 1-D tiling, the corresponding estimated processing times is
\begin{equation}
T = \beta \frac{N}{N_p} \{ \max{(1- 6L_y N_p/N , ~\frac{2SL_y}{\beta B_y} N_p/N)} +6 L_y N_p/N \}.
\label{eq:tot1dov}
\end{equation}
Going to 2-D tiling we need to gather not contiguous data in GPU memory for 
efficient node-to-node communication, which impacts overlap possibilities, 
or work with multiple small and less efficient node-to-node communication steps.  
Details of this will be explained in a later section.
One is then forced to perform a communication step for non-contiguous data first
(in the $Y$ direction in our case), followed by overlapped bulk computation 
and communication of contiguous buffers and finally by computation of border data. 
%The corresponding time estimate is:
%\begin{equation}						 
%T = \beta \frac{N}{N_p} \{ \max{(1 - \frac{6(n_xL_y+n_yL_x)}{N}, %~\frac{SL_y}{n_y \beta B_y} N_p/N) } + \frac{SL_x}{ n_x \beta B_x} N_p/N 
%+ 6(n_xL_y + n_yL_x)/N   \}
%\label{eq:tot2dovGen}
%\end{equation}
The corresponding estimate, that for simplicity we write only for square lattices and for the
not necessarily optimal case $n_x = n_y = \sqrt{N_p}$, is:
\begin{equation}						 
T = \beta \frac{N}{N_p} \{ \max{((1 - 12 \sqrt{N_p/N}), ~\frac{2S}{\beta B_y} \sqrt{N_p/N}) } + \frac{2S}{ \beta B_x} \sqrt{N_p/N} 
+ 12 \sqrt{N_p/N}   \}
\label{eq:tot2dov} 
\end{equation}

Extracting accurate predictions from Eqs. \ref{eq:tot2d} to \ref{eq:tot2dov} 
is made more difficult by the fact that communications bandwidths depend on 
the transfer direction (as already remarked) and {\em also} on buffer size. 
We have performed direct bandwidth measurements for relevant values of the buffer sizes 
(shown later, see \figurename~\ref{fig:psg-bw}, where several important details 
of this measurement are discussed); 
putting those data into our model equations, 
we predict a pattern of scaling violations shown in \figurename~\ref{plot:model} for one typical lattice size. 
Several neglected factors may change the details of our predictions, 
so we stress again that we use our theoretical estimates for guidance only. 
Two main lessons emerge from our models: i) overlapping communication and 
computation has a strong impact on performance; if we do so to the extent 
made possible by system features, one can expect limited violations 
to scaling on reasonably-sized lattices and on a fairly large number of GPUs, 
and, ii) contrary to naive expectation, an 1-D tiling of the lattice may 
have good performances up to a reasonably large number of processors.

Based on this overall picture, we have prepared and tested several 
parallel versions of the code that we describe and compare in the following.

%%%%%%%%%%%%%%%%%%%%%%%%%%%%%%%%%%%%%%%%%%%%%%%%%%%%%%%%%%%%%%%%%%%%%%
\begin{figure}
%\centering
\begin{lstlisting}[basicstyle=\scriptsize]
#ifdef MPI_REGULAR
cudaMemcpy ( sndbuf_h, sndbuf_d, N, cudaMemcpyDeviceToHost );
MPI_Sendrecv(
  sndbuf_h, N, MPI_DATATYPE, nxt, 0,
  rcvbuf_h, N, MPI_DATATYPE, prv, 0,
  MPI_COMM_WORLD; MPI_STATUS_IGNORE );
cudaMemcpy ( rcvbuf_d, rcvbuf_h, N, cudaMemcpyHostToDevice );
#endif

#ifdef CUDA_AWARE_MPI
MPI_Sendrecv(
  sndbuf_d, N, MPI_DATATYPE, nxt, 0,
  rcvbuf_d, N, MPI_DATATYPE, prv, 0,
  MPI_COMM_WORLD; MPI_STATUS_IGNORE );
#endif
\end{lstlisting}
\caption{Definition of two CUDA-codes for a bi-directional memory-copy 
of buffers allocated on two GPUs; the first case is a {\em regular} MPI 
implementation requiring to move explicitly data from GPU to host and 
vice-versa; the latter case uses {\em CUDA-aware} MPI allowing to call  
directly the {\tt MPI\_Sendrecv} with pointers to buffers in GPU memory 
as source and destination parameters.}
\label{cuda-example}
\end{figure}

%%%%%%%%%%%%%%%%%%%%%%%%%%%%%%%%%%%%%%%%%%%%%%%%%%%%%%%%%%%%%%%%%%%%%%

\subsubsection*{Programming Models for Multi-GPU applications}

In this sub-section we briefly overview programming models relevant for 
multi-GPU codes. The goal is to make code development and management 
easier and communications more efficient.  

Large GPU clusters are widely heterogeneous computing systems: compute nodes 
have one or (usually) more CPUs; each CPU acts as host for a variable number 
of GPUs, ranging typically from 1 to 4; in the cluster that we have used for our 
tests each node has 2 CPUs and each CPU hosts 4 GPUs. GPUs are directly connected 
to their host through a PCIe interface, which has reasonably high bandwidth 
(several Gbytes/sec) but also long startup latency ($\ge 1 \mu$sec). 
The network interface is also connected to one of the CPUs via PCIe.
The complexity of this structure implies that what, at the application level, 
is a plain GPU-to-GPU communication may involve different routes, different 
communication strategies, and correspondingly different performances. 
We discuss here two key aspects of the problem, namely i) a programming 
environment able to specify in a unified way all different communication 
patterns and, ii) the ensemble of run-time support features that help 
maximize effective communication bandwidth for any possible pair of 
communication end-points.

Concerning the first point, a reasonable approach is to use the well known MPI
communication libraries that currently also support GPUs; 
we then associate one MPI process to each GPU, so MPI libraries are able to 
automatically handle the transfer of data buffers from GPU to GPU in the most 
appropriate way. 
Transferred buffers must be allocated at contiguous locations on memory; however, 
transfers of non-contiguous buffers can also be handled automatically by MPI, 
using the derived {\tt vector} data type: 
the {\tt vector} data type describes how data buffers are placed in memory and the 
library automatically packs data into a contiguous buffer, perform the MPI communication 
and then unpack data at destination. 
Note that in regular MPI versions -- i.e. without GPU support -- these buffers had 
to be allocated on the host CPU, so each data transfer has to be preceded and followed 
by an explicit data move from/to GPU and its host. 
CUDA-aware MPI~\cite{cuda-aware_mpi} improves on this, allowing to specify
buffers allocated on the GPU memory as arguments of the MPI operations, making 
codes terser and more readable; \figurename~\ref{cuda-example} compares the CUDA 
definitions of a function that performs a bi-directional remote memory-copy of a 
buffer allocated on the memory of two GPUs, using regular MPI and CUDA-aware MPI.

Armed with a clean way to specify the communication patterns needed by 
our program, we must now make sure that all possible steps are taken to reduce 
the latency of each communication, as this has a critical impact on 
the scaling performance of the complete code. This is done by enabling a variety 
of features, available in the low-level communication libraries; here we describe 
the most relevant points. 

For GPUs attached to the same host-CPU, CUDA-IPC moves data directly across GPUs
without staging on CPU-memory. This makes communication faster~\cite{sbac-pad13}.
GPUs attached to different CPUs of the same node communicate through 
CPU-memory staging; here pipelining helps to shorten communication latency.
For GPUs belonging to different nodes GPUDirect RDMA moves short data packets from the 
GPU to the network interface without any involvement of the host CPU. For longer  
data packets due to PCIe architectural bottlenecks, RDMA becomes 
less effective, see~\cite{rdma}. 
In this case, GPUDirect simplifies the operation by sharing a common staging 
region between the GPU and the network interface.

\subsubsection*{1-D Splitting}

\begin{figure}
\centering
\includegraphics[width=0.9\textwidth]{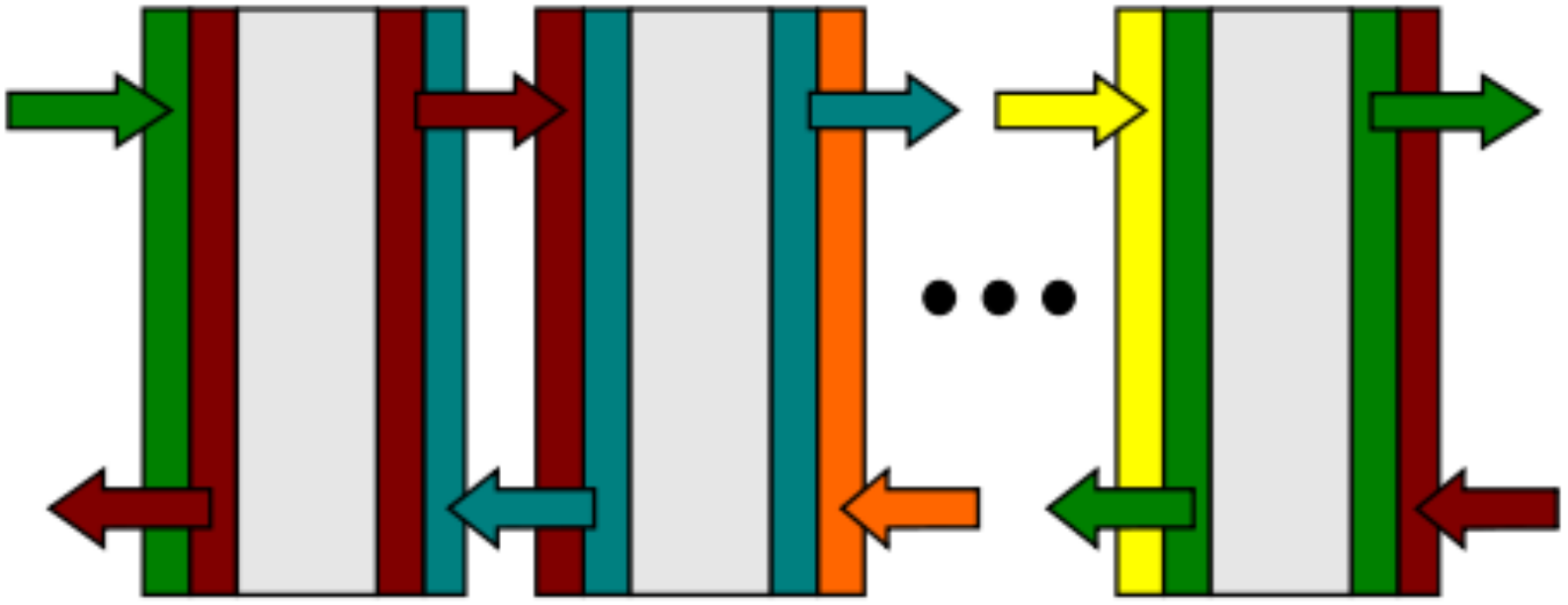}
\caption{\label{1D-tiling} 1-D tiling of a lattice on $N_p$ GPUs virtually 
ordered along a ring.}
\end{figure}

In this case, we divide a lattice of size $L_x \times L_y$ 
on $N_p$ GPUs tiling along just one dimension. In our case, since lattice is allocated 
by column-major order, we split the lattice along the $X$ dimension and 
then each GPU allocates a {\em sub-lattice} of size $ L_x / N_p \times L_y$, 
see \figurename~\ref{1D-tiling}.
%
%%%%%%%%%%%%%%%%%%%%%%%%%%%%%%%%%%%%%%%%%%%%%%%%%%%%%%%%%%%%%%%%%%%%%%%%
%
%
%%%%%%%%%%%%%%%%%%%%%%%%%%%%%%%%%%%%%%%%%%%%%%%%%%%%%%%%%%%%%%%%%%%%%%%%
%
This 1D tiling implies a virtual ordering of the GPUs along a ring, so each GPU 
is connected with a previous and a next GPU; at the beginning of each time-step, 
GPUs must exchange data, since cells close to the right and left edges 
of the sub-lattice of each GPU needs data allocated on the logically previous 
and next GPUs, see again \figurename~\ref{1D-tiling}.

%%%%%%%%%%%%%%%%%%%%%%%%%%%%%%%%%%%%%%%%%%%%%%%%%%%%%%%%%%%%%%%%%%%%%%%%
%
\begin{figure}
\centering
\begin{lstlisting}[basicstyle=\scriptsize]
// Computing propagate over lattice bulk
prop_Bulk     <<< dimGridB, dimBlockB, 0, stream[0] >>> ( ... );
bc_Bulk       <<< dimGridB, dimBlockB, 0, stream[0] >>> ( ... );
collide_Bulk  <<< dimGridB, dimBlockB, 0, stream[0] >>> ( ... );
// Update halos
pbc_c();
// Computing propagate on left columns
prop_L     <<< dimGridLR, dimBlockLR, 0, stream[1] >>> ( ... );
collide_L  <<< dimGridB, dimBlockB, 0, stream[1] >>> ( ... );
// Computing propagate on right columns
prop_R     <<< dimGridLR, dimBlockLR, 0, stream[2] >>> ( ... );
collide_R  <<< dimGridLR, dimBlockLR, 0, stream[2] >>> ( ... );
\end{lstlisting}
\vspace*{2.5mm}
\includegraphics[width=\textwidth]{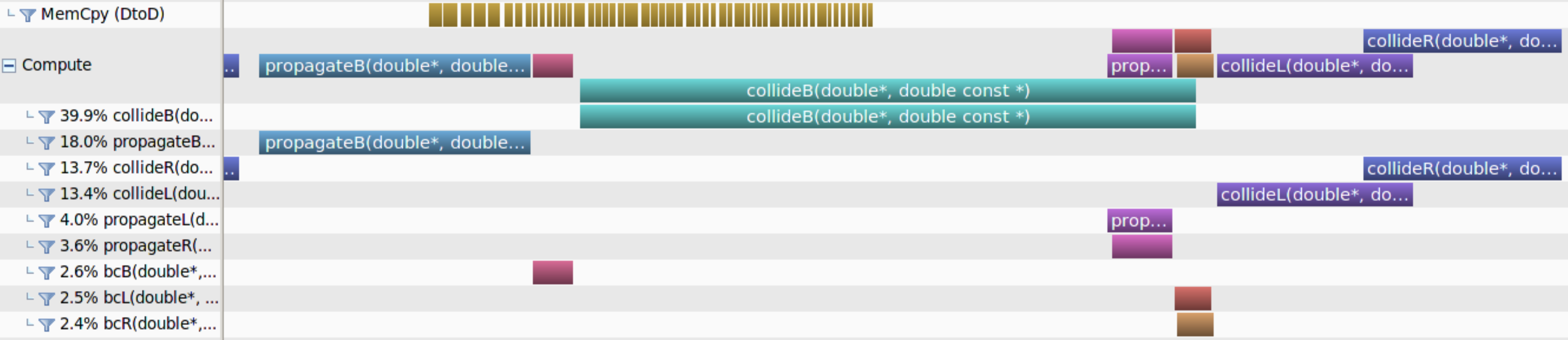}
\caption{Sample code and timeline of a multi-GPU code executed by 
each MPI-process using the 1-d tiling of the lattice. 
Communications are performed by {\tt pbc} and use CUDA-aware 
MPI; this step is then translated into CUDA device-to-device memory copies 
since this example refers to GPUs allocated on the same host; 
communication overlaps with the execution of the {\tt propagate}, 
{\tt bc} and {\tt collide} kernels on the bulk of the lattice. 
After MPI communications have  
completed, the computational kernels acting on the right and left 
edges of the lattice can start; they do so as soon as GPU resources 
become available.}
\label{flowchart}
\end{figure}
%%%%%%%%%%%%%%%%%%%%%%%%%%%%%%%%%%%%%%%%%%%%%%%%%%%%%%%%%%%%%%%%%%%%%%%%%%

For processing, the lattice is divided in three regions: two regions of 
size $3\times Ly$ include the three leftmost and the three rightmost column-borders, 
while another region includes the central part of the lattice that we call the bulk. 
Processing the left and right regions can start only after the left 
and right halos have been updated, while processing on the bulk can start 
immediately and overlaps with the update of halos.
Each MPI process executes a code structured as in \figurename~\ref{flowchart} top: 
it runs a CUDA-stream executing in sequence the {\tt propagate}, {\tt bc} 
and {\tt collide} kernels on the bulk region. 
In parallel, the host-PC executes the {\tt pbc\_c} ({\tt \_c} stands for contiguous) 
function which performs MPI communications to update left and right halos with 
neighbor GPUs in the ring. 
After all data transfers are complete, two additional CUDA-streams 
start {\tt propagate}, {\tt bc} and {\tt collide} 
on the left and right border regions.

\figurename~\ref{flowchart} bottom shows the timeline execution 
of the code.
We directly see that an efficient implementation of {\tt pbc} helps to 
enlarge the region in which linear scaling is possible; as we partition the lattice 
onto a larger and larger number of processors, the combined execution times 
of {\tt propagate}, {\tt bc} and {\tt collide} 
reduces accordingly, while the execution time of {\tt pbc} remains 
approximately constant. 
Eventually, {\tt pbc} takes longer than the 
computational kernels, and scaling violations occur. 
There is no way to escape this situation asymptotically, but an efficient 
implementation of {\tt pbc\_c} delays the onset of scaling violations.
We have found that the implementation of {\tt pbc\_c} through 
a sequence of CUDA-aware MPI operations gives good results; in our case,
26 populations must be moved for each boundary site,
corresponding to 52 MPI operations.
If the lattice is large enough in the Y direction ($\ge 512$ points)
the overheads associated to separate MPI operations are negligible. For 
smaller lattices it may be useful to 
pack data in a contiguous buffer and perform just one larger MPI transfer. 
We discuss this further optimization in the next section, where we also consider
the fusion of {\tt propagate} and 
{\tt collide} into just one CUDA kernel.

\subsubsection*{2D Splitting}

%%%%%%%%%%%%%%%%%%%%%%%%%%%%%%%%%%%%%%%%%%%%%%%%%%%%%%%%%%%%%%%%%%%%%%%%
%
\begin{figure}
\centering
\includegraphics[width=\textwidth]{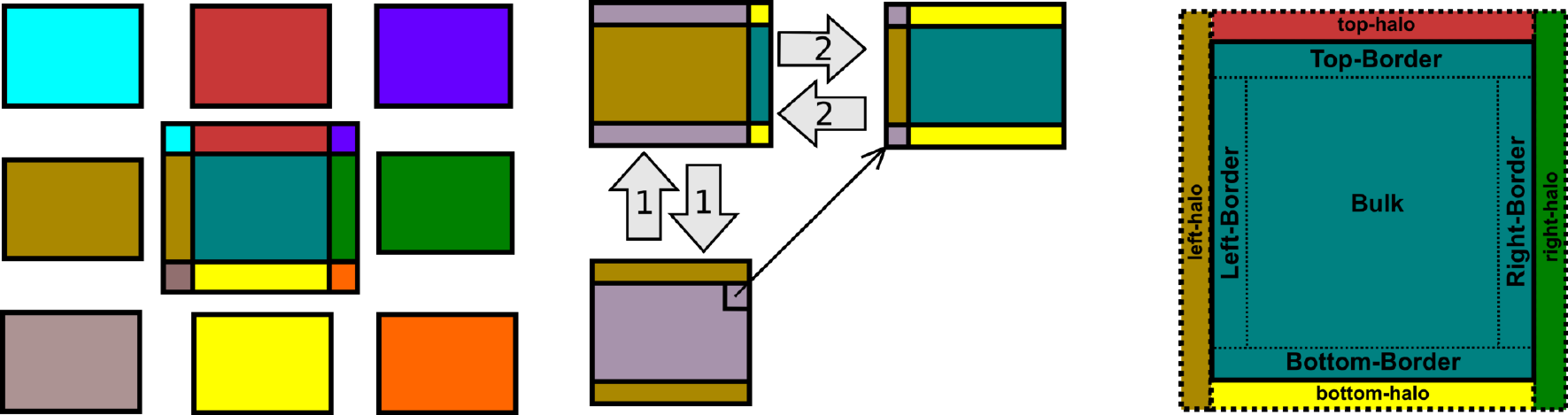}
\caption{\label{2D-tiling} 
2-D tiling of the lattice on $N_p$ GPUs. Left: diagram 
of the tiles and of the corresponding halos. Center: communication patterns 
to update halos belonging to a given tile. Righ: halo regions surrounding the tile.}
\end{figure}
%
%%%%%%%%%%%%%%%%%%%%%%%%%%%%%%%%%%%%%%%%%%%%%%%%%%%%%%%%%%%%%%%%%%%%%%%%

Code organization using a 2-D tiling is slightly more complex.
We split the lattice on a grid $n_x \times n_y$  GPUs, virtually arranged 
at the edges of a 2D mesh.
Each GPU needs data  allocated on eight neighbor GPUs, 
see~\figurename~\ref{2D-tiling}, left.

All needed data transfers (from adjacent and from diagonal neighbor nodes) 
can be done by performing a sequence (see again ~\figurename~\ref{2D-tiling}, center and right) 
in which first all nodes exchange data along one of the two directions, not including
halo elements; when this is completed a further step in the orthogonal direction is started, 
including this time also halo elements.

One of the two communications steps (the one in the Y+ and Y- directions in our case) 
implies non-contiguous data elements. As discussed in an earlier section,
communications of non contiguous buffers is automatically handled by MPI 
using the {\tt vector} derived data type. The corresponding standard library 
gathers all data elements into an intermediate buffer, starts a transfer 
operation for the intermediate buffer and finally scatters received data 
items to their final destination. 
We tested two well-known CUDA-aware MPI libraries, OpenMPI and MVAPICH2. 
Results were unsatisfactory for two reasons: 
i) OpenMPI is affected by high overheads because of the many calls to copy 
all the pieces of data into the intermediate MPI buffer on the host;
ii) MVAPICH2 do not use {\em persistent} intermediate MPI buffers, 
that are allocated and de-allocated on the GPU at each time step; 
the corresponding overhead in doing that is in our case too large, 
and it can be easily avoided using persistent allocation of communication 
buffers on GPU memory~\footnote{
We provided these as feedbacks to OpenMPI and MVAPICH2 development communities;  
for both MPI implementations improvements for the mentioned issues are 
planned for future releases.}. 

We have overcome these issues developing a custom communication library 
that uses persistent send and receive buffers, allocated once on the GPUs 
at program initialization. 
Every time a communication of non contiguous buffers is needed, function {\tt pbc\_nc}
({\tt \_nc} stands for non-contiguous) starts the {\tt pack} kernel 
to gather non-contiguous data into a contiguous buffer allocated on the GPU.
When this is done, an MPI communication is started, followed by a final 
scatter of the received data.
\figurename~\ref{fig:pack-unpack} in the appendix shows a simplified CUDA 
implementation of the {\tt pack} and {\tt unpack} kernels, and 
\figurename~\ref{fig:nc-buffers} shows a sample code to handle 
data transfers among non contiguous buffers. 

This strategy has the advantage that, for each halo update, only one MPI 
communication is needed. 
This avoids overheads associated to start MPI transfers, and  keeps the size 
of MPI buffers large enough to minimize overheads caused by CUDA-IPC set-up. 
The advantages of this approach are relevant also for updating contiguous halos; 
for this reason we adopt it for communications in both directions. 

%%%%%%%%%%%%%%%%%%%%%%%%%%%%%%%%%%%%%%%%%%%%%%%%%%%%%%%%%%%%%%%%%%%%%%
\begin{figure}[t]
\centering
\includegraphics[width=\textwidth]{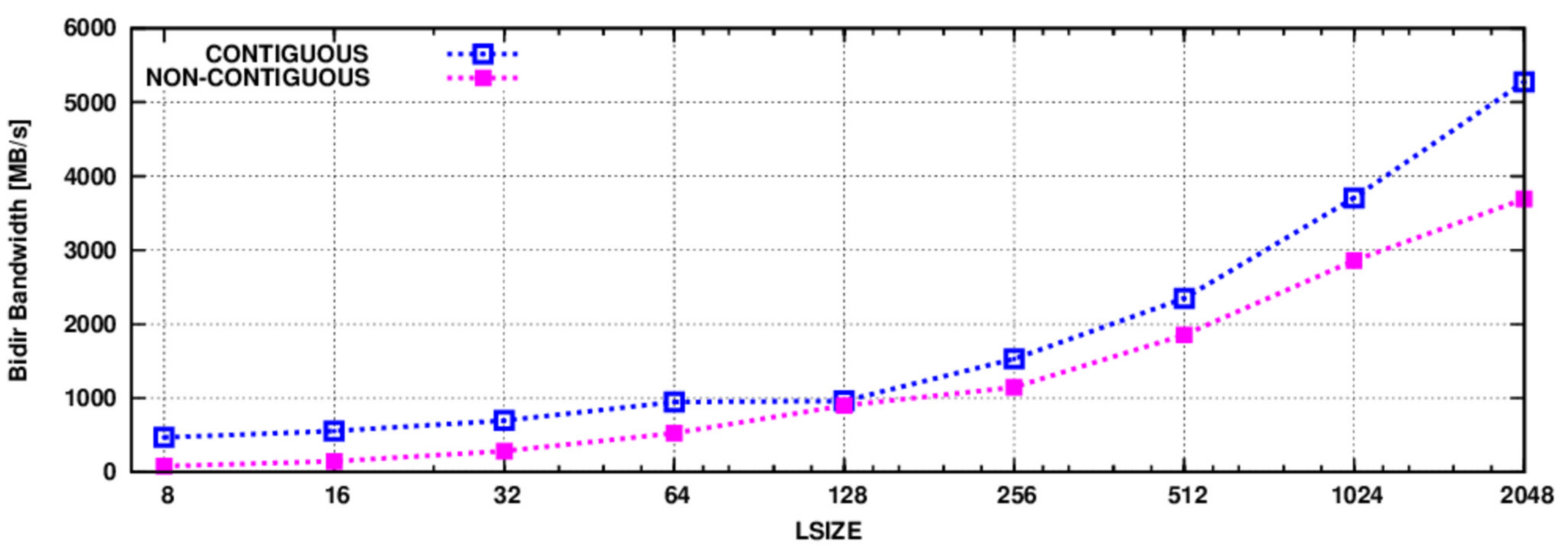}
\caption{
\label{fig:psg-bw}
Measured effective bi-directional bandwidth to update 
contiguous and non contiguous halos as a function of the 
size of the lattice tile. The test has been done on two K80 
systems on two remote nodes, interconnected by an Infiniband 
FDR network with GPUDirect RDMA enabled.
}
\end{figure}
%
%%%%%%%%%%%%%%%%%%%%%%%%%%%%%%%%%%%%%%%%%%%%%%%%%%%%%%%%%%%%%%%%%%%%%%%%

\figurename~\ref{fig:psg-bw} reflects the global result of this optimization 
effort, showing the effective bi-directional bandwidth measured in the update 
of memory-contiguous and non contiguous halos as a function of the corresponding 
lattice size. 
This test involves two K80 boards attached to two different host-CPUs 
interconnected through Infiniband network.
We see that, as expected, non-contiguous halos have a reduced effective 
bandwidth, but the difference between the two cases is not too large. 
The data shown in \figurename~\ref{fig:psg-bw} has been used in the 
scaling prediction models that we have discussed before.

%%%%%%%%%%%%%%%%%%%%%%%%%%%%%%%%%%%%%%%%%%%%%%%%%%%%%%%%%%%%%%%%%%%%%%%%
%
\begin{figure}
\centering
\begin{lstlisting}[basicstyle=\scriptsize]
// Update non-contiguous halos
pbc_nc()

// Computing over bulk
prop_Bulk    <<< .dimGridB, dimBlockB, 0, stream[0] >>> ( ... )
collide_Bulk  <<< dimGridB, dimBlockB, 0, stream[0] >>> ( ... );

// Update contiguous halos
pbc_c()

// Computing propagate on left columns of the lattice
prop_L     <<< dimGridLR, dimBlockLR, 0, stream[1] >>> ( ... );
collide_L  <<< dimGridLR, dimBlockLR, 0, stream[1] >>> ( ... );

// Computing propagate on right columns of the lattice
prop_R    <<< dimGridLR, dimBlockLR, 0, stream[2] >>> ( ... );
collide_R <<< dimGridLR, dimBlockLR, 0, stream[2] >>> ( ... );

// Computing propagate on top rows of the lattice
prop_T    <<< dimGridTB, dimBlockTB, 0, stream[3] >>> ( ... );
bc_T      <<< dimGridTB, dimBlockTB, 0, stream[3] >>> ( ... );
collide_T <<< dimGridTB, dimBlockTB, 0, stream[3] >>> ( ... );

// Computing propagate on bottom rows of the lattice
prop_B    <<< dimGridTB, dimBlockTB, 0, stream[4] >>> ( ... );
bc_B      <<< dimGridTB, dimBlockTB, 0, stream[4] >>> ( ... );
collide_B <<< dimGridTB, dimBlockTB, 0, stream[4] >>> ( ... );

\end{lstlisting}
\caption{\label{2D-code-with-overlap} 
Scheduling of operations for the code using 2-D tiles.
}
\end{figure}
%
%%%%%%%%%%%%%%%%%%%%%%%%%%%%%%%%%%%%%%%%%%%%%%%%%%%%%%%%%%%%%%%%%%%%%%%%

For the processing steps of the algorithm, the lattice is divided in five 
regions, see~\figurename{\ref{2D-tiling}}, right: 
two regions of $3 Ly$ sites, including the three leftmost 
and the three rightmost columns, two regions of size $3 L_x$ including 
the three topmost and lowermost rows and the central part of the lattice 
including all bulk sites. 
The code in \figurename~\ref{2D-code-with-overlap} shows how we schedule 
operations. 
We first exchange the (non-contiguous) top and bottom halos; when 
this operation has completed, we start processing the lattice bulk on a 
GPU stream, and in parallel we update the contiguous left and right halos.
After all halos have been updated, we start separate GPU streams 
processing the left, right, top and bottom borders.

%%%%%%%%%%%%%%%%%%%%%%%%%%%%%%%%%%%%%%%%%%%%%%%%%%%%%%%%%%%%%%%%%%%%%%%%
%
\begin{figure}
\includegraphics[width=\textwidth]{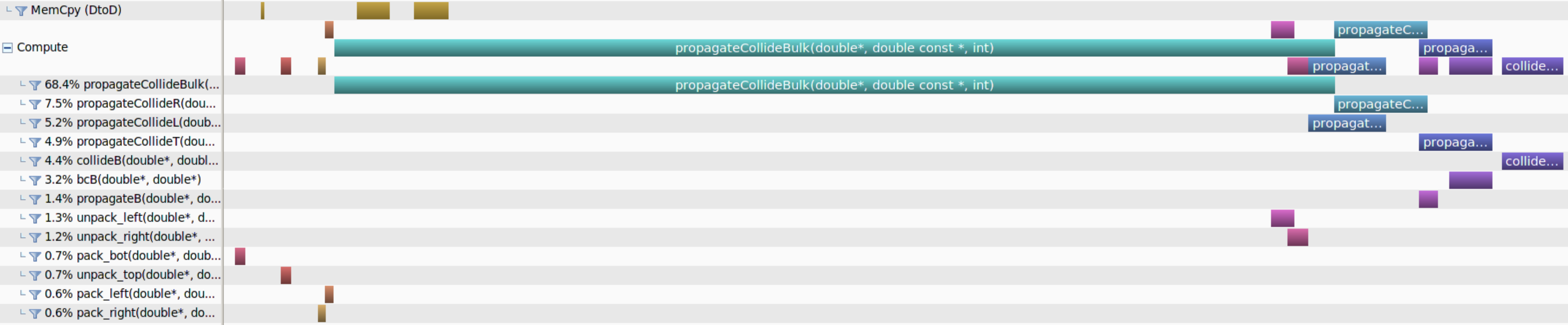}
\caption{\label{2D-code-scheduling} 
Execution timeline of the code using a 2-d tiling, as shown by the NVIDIA profiler. 
In this example, we first execute update of non-contiguous halos starting  
first the {\tt pack\_bot} kernel, and after MPI communication the {\tt unpack\_tot}. 
These operations can not overlap with other data-processing. After we start 
the {\tt pack\_left} and  {\tt pack\_right} to pack data to update contiguous 
halos, and in parallel we execute the {\tt propagateCollideBulk} kernel to 
process all sites belonging to lattice bulk. As soon as {\tt propagateCollideBulk} kernel 
frees GPU resources, {\tt unpack\_left} and {\tt unpack\_right} kernels 
are executed to update contiguous halos, followed by the processing of 
the 4 border regions.
}
\end{figure}

%%%%%%%%%%%%%%%%%%%%%%%%%%%%%%%%%%%%%%%%%%%%%%%%%%%%%%%%%%%%%%%%%%%%%%%%

With this scheduling it is easy to merge {\tt propagate} and {\tt collide}
for all points on which {\tt bc} kernel does not apply, belonging to the bulk, 
left and right regions. 
On the other hand, top and bottom borders must be processed by a sequence 
of {\tt propagate}, {\tt bc} and {\tt collide} kernels. 
\figurename~\ref{fig:2D-code-with-overlap-fused} shows the final organization 
of the code including these improvements, and \figurename~\ref{2D-code-scheduling} 
shows the corresponding execution timeline as recorded by the NVIDIA profiler.
The update of non-contiguous halos can not overlap (see caption for details) 
with other data-processing operation, because: 
i) MPI communications along Y direction needs to be done before starting 
that along the X direction to update also halos with data coming from 
diagonal neighbor sites, 
ii) and the corresponding {\tt pack} and {\tt unpack} kernels needs to be 
executed before GPU resources become busy in processing the sites of the lattice bulk.   
On the other hand, the update of contiguous halos fully overlaps with 
processing of the bulk region. 
Finally, {\tt unpack} of received data for contiguous halos, and the processing 
of the 4 border regions starts as soon as GPU resources are freed by the kernel 
processing the bulk regions.

%%%%%%%%%%%%%%%%%%%%%%%%%%%%%%%%%%%%%%%%%%%%%%%%%%%%%%%%%%%%%%%%%%%%%%%%
\begin{table}[t]
\caption{
Performance of our full production-ready code, measured  on a lattice 
of $1024 \times 8192$ sites. We show the execution time of each phase of the 
code, the performance of the {\tt propagate} and {\tt collide} kernels, 
the effective performance of the complete code (Global $P$), and 
the MLUPS (Million Lattice Updates per second) metric.
The clock frequencies for the SMs of the K40 and K80 boards are 
set at the ``boosted'' values of $875$ MHz.
}
\label{single-host-perf}
\centering
\resizebox{\textwidth}{!}{
\begin{tabular}{lrrrrrr}
\toprule
                        &       GF100 &  2$\times$GF100  & GK110B    & 2$\times$GK110B     & GK210    & 2$\times$GK210  \\
\midrule
%$T_{\mbox{pbc}}$  (ms) &        0.3  &       1.0        & 0.7       & 0.9                 &  9.4     &  18.8      \\
$T_{\mbox{prop}}$ (ms)  &       60.6  &      30.9        & 25.8      & 12.2                & 32.3     &  19.0      \\
$T_{\mbox{bc}}$   (ms)  &        6.5  &       3.6        & 2.8       & 1.4                 &  1.4     &   0.8      \\
$T_{\mbox{coll}}$ (ms)  &      276.0  &     158.0        & 78.0      & 39.0                & 71.1     &  38.1      \\
\midrule
Propagate (GB/s)        &        81   &     155          & 187       &  376                & 155      &   261      \\
Collide (GF/s)          &       197   &     344          & 696       & 1388                & 764      &  1544      \\
\midrule
%Wallclock time (ms)    &       343.4 &     193.5        & 107.4     & 53.8                & 115      &  54.9      \\
Global $P$. (GF/s)      &       158   &     281          & 506       & 1010                & 519      &  988       \\
MLUPS                   &       24    &     43           & 78        & 156                 &  80      &  153       \\
\bottomrule
\end{tabular}
}
\end{table}
%%%%%%%%%%%%%%%%%%%%%%%%%%%%%%%%%%%%%%%%%%%%%%%%%%%%%%%%%%%%%%%%%%%%%%

%%%%%%%%%%%%%%%%%%%%%%%%%%%%%%%%%%%%%%%%%%%%%%%%%%%%%%%%%%%%%%%%%%%%%%%%

\section{Results Analysis}

In this section we present results for our full production-grade codes, 
that consistently use all paths to performance discussed 
in the previous sections. 

\subsubsection*{One GPU}

We first examine results for just one (or two) GPUs: \tablename~\ref{single-host-perf} 
collects performance results of the full production-ready code running on one host with 
one or two GPUs on a lattice of $1024 \times 8192$ points; the main
computational load is associated to the {\tt propagate} and {\tt collide}
kernels, as expected. 
Memory bandwidth (relevant for {\tt propagate}) is close to $55\%$ of 
the theoretical peak for the C2050 accelerator; it reaches $\approx 65\%$ 
of peak for the K40 and the K80.
The {\em Kepler} processor is more efficient from the point of view of floating-point 
throughput, as measured by the FP performance of the {\tt collide} 
kernel, reaching $\approx 43\%$ versus $\approx 38\%$ for the C2050 board; the 
K80 board exploits its larger register 
file and shared memory to reach $\approx 53\%$ of peak.
On a dual-K40 system and on a K80 board using both GPUs, the {\tt collide} kernel 
largely breaks the sustained double precision $1$ Tflops performance barrier;
also the global performance figures of the full code, which take into account all 
execution phases, are satisfactory: we measure an efficiency of respectively 
$\approx 31\%$ and $\approx 34\%$ of the raw peak floating-point throughput.  

\subsubsection*{Performance Comparison}

It is interesting to compare the performance delivered by GPUs with that 
of other recent processors; this is done in \tablename~\ref{comparison}, 
where we compare performance figures on GPUs systems with those of implementations 
of the same code developed and optimized for multi- and many-core 
Intel systems. 
We consider a dual-E5-2630 V3 system, with two eight-core Haswell V3 processors, 
and a 61-core Xeon-Phi processor, the latest accelerator based on the 
Intel MIC many-core architecture. For these architectures we have optimized 
the code parallelizing the execution over all available cores and 
using SIMD instructions within the cores; for details, see \cite{caf13,ccp12} and \cite{iccs13,chep13}. 
Performances of the {\tt propagate} and 
{\tt collide} kernels are significantly faster on the the K40 and K80 boards than on the other systems:
the {\tt propagate} kernel is approximately $2-3X$ faster than the dual-CPU systems,  
and $1.9X$ and $2.7X$ faster than the Xeon-Phi; 
this is also true for {\tt collide}, where GPUs are $3.7X$ and $7.6X$ 
faster than CPUs, and $2.0X$ and $4.0X$ faster than the Xeon-Phi.

\subsubsection*{Energy Efficiency}

\tablename~\ref{comparison} also shows data on energy efficiency 
(that we normalize to the average energy needed to process one lattice site). We estimate this quantity using
published data on the {\em Thermal Design Point} (TDP), an upper bound of the 
power consumption of the processors; this gives only a rough estimate of the 
energy efficiency, also because all other sources of power consumption (host, memories, devices, \ldots) are neglected. 
Taking these reservations into account, GPUs are more energy-efficient than the other
processors: for instance, the K80 system is $\approx 7X$ better than the
dual-CPU system and $\approx 4X$ better than the Xeon-Phi. When considering 
these results one must keep in mind that co-processors (GPUs and Xeon-Phi) 
operates with the support of a host processor: even if the latter is little 
used during the computation, it still draws an amount of power that is not 
necessarily a small fraction of the total energy budget.

%%%%%%%%%%%%%%%%%%%%%%%%%%%%%%%%%%%%%%%%%%%%%%%%%%%%%%%%%%%%%%%%%%%%%%
\begin{table}
\centering
\caption{
\label{comparison}
Performance comparison of the {\tt propagate} and {\tt collide} kernels running 
on several systems: a dual eight-core E5-2630 2.4 GHz (Intel Haswell V3) processor, 
an Intel Xeon-Phi accelerator and a K40 and K80 board. 
$\epsilon$ is the effective performance w.r.t. peak performance.
For {\tt collide} we also measure performance in {\em Million Lattice Updates
per Second} (MLUPS). Finally we list the energy needed
to update one lattice cell, estimating the power used by processors from published 
data on their Thermal Design Point (TDP).
}
\resizebox{\textwidth}{!}{
\begin{tabular} {lrrrr}
\toprule
                  &  dual E5-2630 v3     & Xeon-Phi 7120X       & Tesla K40  & Tesla K80 \\
\midrule
propagate (GB/s)  &  88                  & 98                   & 187        & 261   \\       
$\epsilon$        &  75\%                & 28\%                 & 65\%       & 54\%  \\
\midrule
collide (GF/s)    & 222                  & 362                  & 696        & 1544  \\
$\epsilon$        &  36\%                & 30\%                 & 42\%       & 53\%  \\
MLUPS             &  29                  & 54                   & 107        & 220   \\
\midrule
TDP (Watt)        & 2$\times$85          & 300                  & 235        & 300   \\
Energy ($\mu J$/site)    & 7.3           & 5.5                  & 2.5        & 1.2   \\
\bottomrule
%
%% (*) I dati di "dual E5-2630 v3" sono estratti dal paper di MARINE15 e si 
%% riferiscono ad un lattice di dimensione 4800x4096
%
\end{tabular}
}
%{(*)\scriptsize using the max TDP declared by vendors}
%}
\end{table}
%%%%%%%%%%%%%%%%%%%%%%%%%%%%%%%%%%%%%%%%%%%%%%%%%%%%%%%%%%%%%%%%%%%%%%

%%%%%%%%%%%%%%%%%%%%%%%%%%%%%%%%%%%%%%%%%%%%%%%%%%%%%%%%%%%%%%%%%%%%%%
\begin{figure}[t]
\centering
\includegraphics[width=\textwidth]{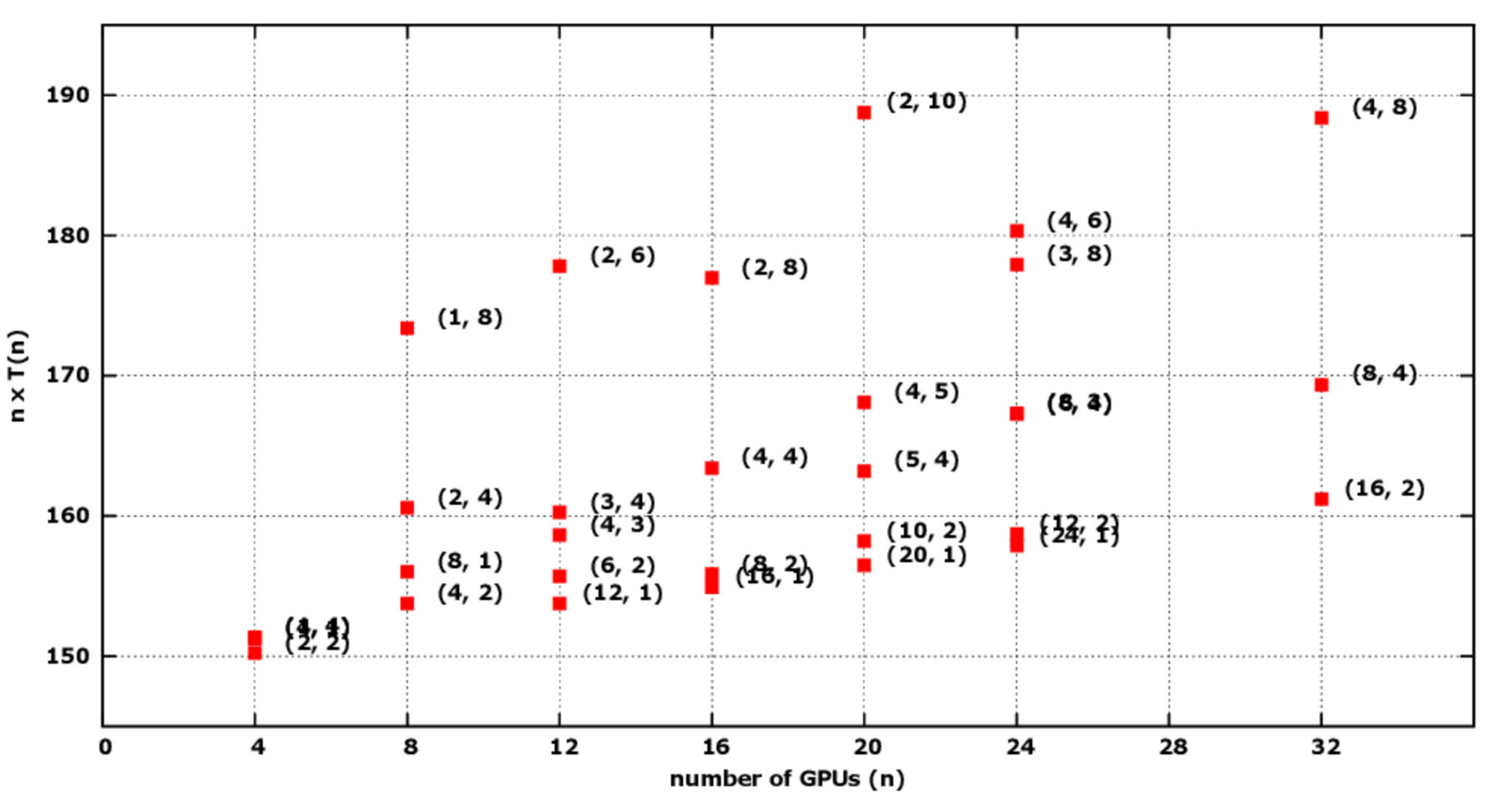}
\caption{Benchmark results on a lattice of $3600 \times 3600$ sites.
For a varying number of GPUs ($n$)  we plot (in arbitrary units) $n \times T(n)$
for all possible 1-D and 2-D decompositions of the lattice (
$T(n)$ is the wall-clock execution time); this product stays constant 
if the code enjoys perfect scalability.}
\label{fig:dataBase}
\end{figure}
%%%%%%%%%%%%%%%%%%%%%%%%%%%%%%%%%%%%%%%%%%%%%%%%%%%%%%%%%%%%%%%%%%%%%%

\subsubsection*{Multi-GPU}

We now move to consider scaling results for multi-GPU codes. 
Following our introductory discussion, we expect that -- contrary to expectation -- 
an 1D tiling of the lattice may be as efficient or even more efficient than 
a 2D tiling up to relatively large number of GPUs. 
We settle the question experimentally, measuring the performance 
of the codes described in the previous sections on several medium-size 
to large lattices, using all possible tilings consistent with the number of 
available GPUs. 
Our tests have been run on a GPU cluster installed at the NVIDIA Technology Center. 
Each node is a dual socket 10-core Ivy Bridge E5-2690 v2 at 3.00GHz, with 
4 K80 GPUs.  
Nodes are interconnected with an Infiniband FDR network, and 
up to 32 GPUs are available. 

%%%%%%%%%%%%%%%%%%%%%%%%%%%%%%%%%%%%%%%%%%%%%%%%%%%%%%%%%%%%%%%%%%%%%%
\begin{figure}[t]
\centering
\includegraphics[width=\textwidth]{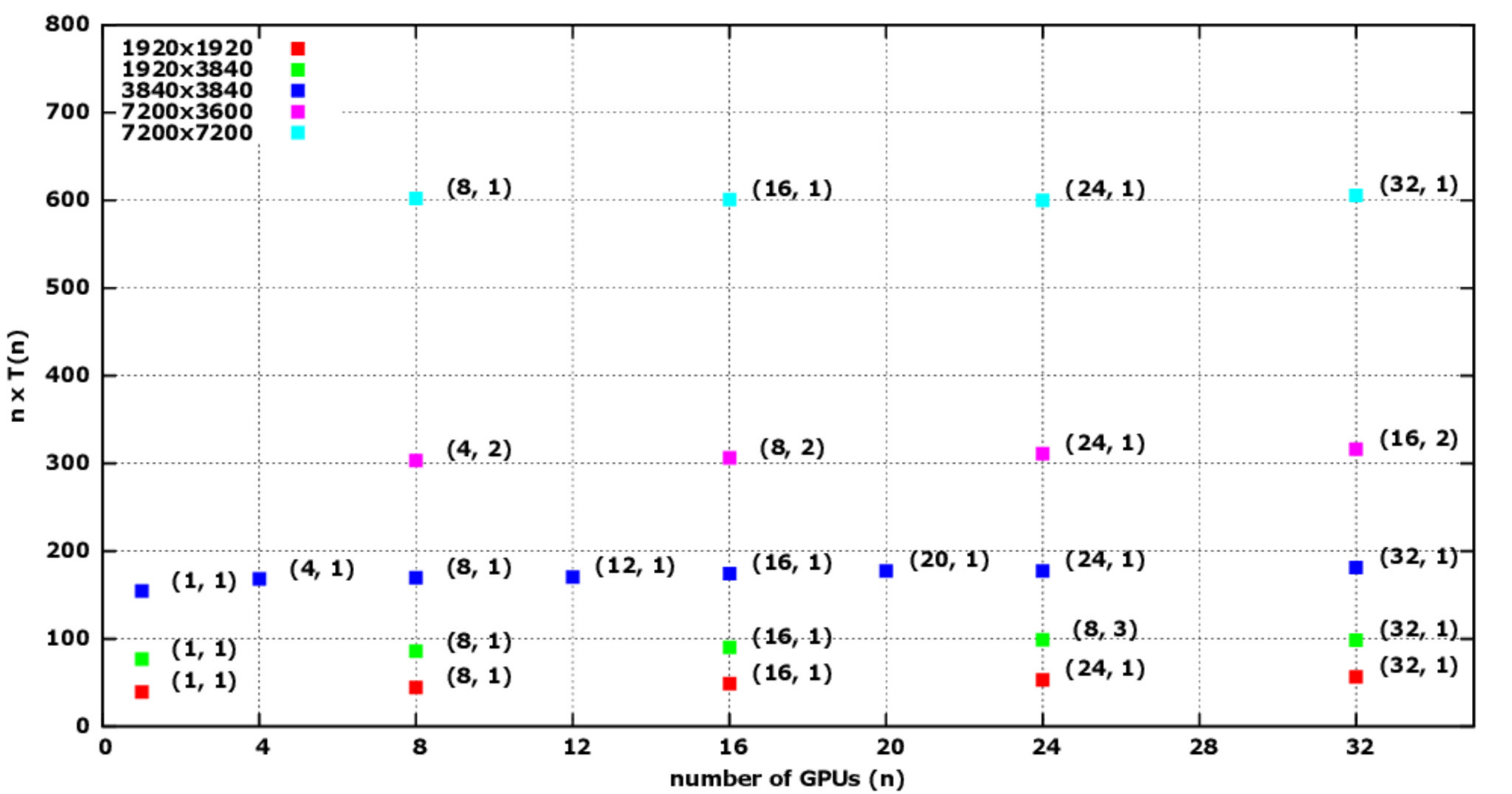}
\caption{Measured values (in arbitrary units) of $n \times T(n)$   
for the more efficient decomposition of the lattice, for several lattice sizes
and for several numbers of GPUs ($n$); 
this product stays constant if the code enjoys perfect scalability.}
\label{fig:best}
\end{figure}
%%%%%%%%%%%%%%%%%%%%%%%%%%%%%%%%%%%%%%%%%%%%%%%%%%%%%%%%%%%%%%%%%%%%%%

\figurename~\ref{fig:dataBase} presents a sample subset of our results,  
plotting  $n \times T(n)$  (in arbitrary units) on a lattice of 
$3600 \times 3600$ points for almost all possible 1-D and 2-D tilings 
of the lattice on $n$ GPUs. 
This quantity is constant if the program enjoys perfect scaling, so it is 
a direct measurement of scaling violations. 
Scaling violations are less than 10\% up to 32 GPUs, and 
some tiling choices are clearly more efficient then others; 
as predicted by our model, the 1D tiling enjoys good 
scaling up to a reasonably large (24 in this case) number of GPUs.
From this data (and from equivalent data for other lattice sizes) and 
for each value of $n$, we derive the best tiling choice; this is shown in 
\figurename~\ref{fig:best}, that contains results for all lattice sizes 
that we have considered. 
We see here that scaling violations are relatively small (and of course 
smaller for larger lattices) on all lattices and for all $n$ that we have 
tested.

%%%%%%%%%%%%%%%%%%%%%%%%%%%%%%%%%%%%%%%%%%%%%%%%%%%%%%%%%%%%%%%%%%%%%%
\begin{figure}[t]
\centering
\includegraphics[width=\textwidth]{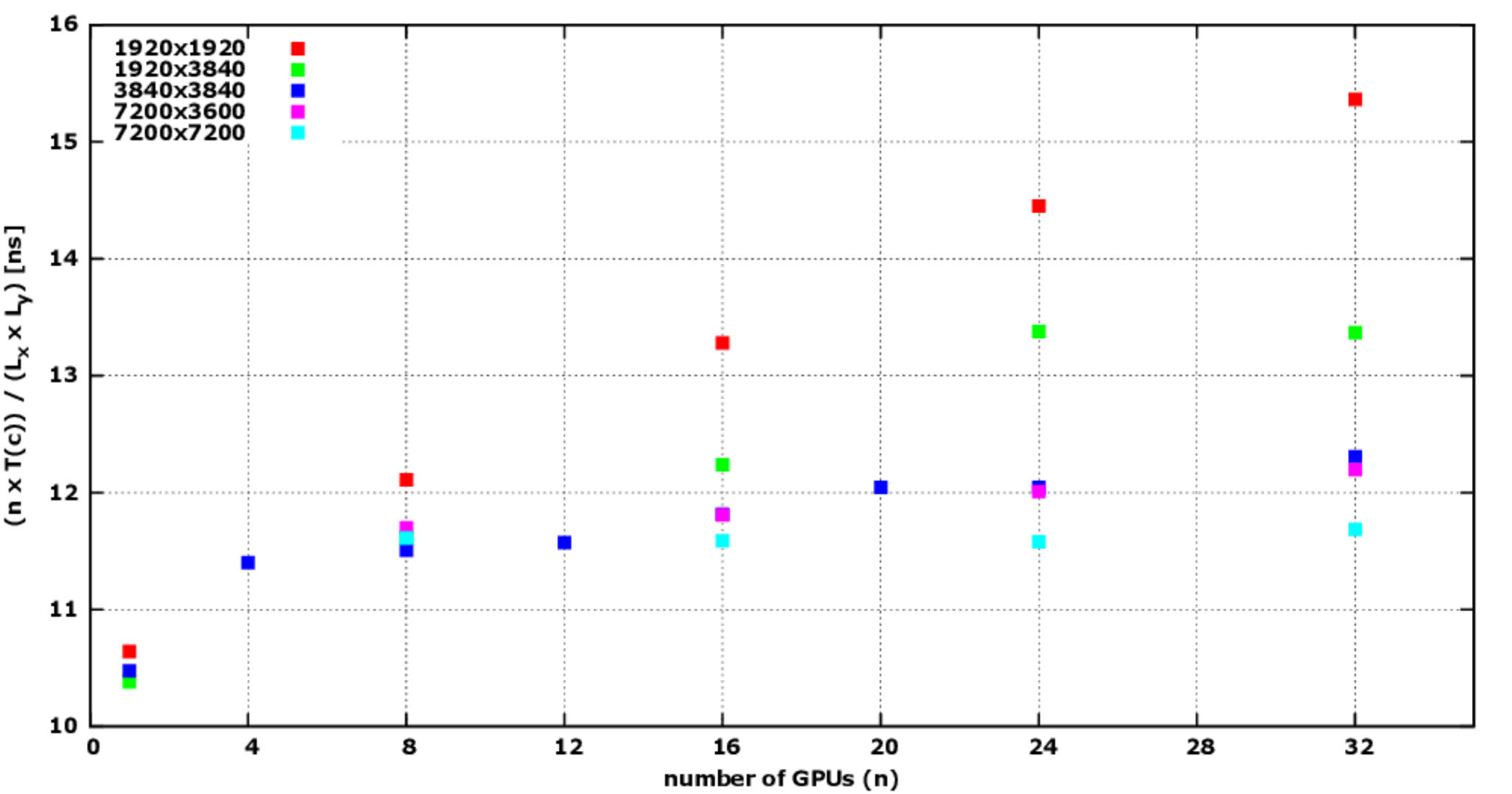}
\caption{Measured values (in nsec) of the average time needed by one GPU 
to process one lattice site as a function of the number of GPUs ($n$), for the 
more efficient decomposition of the lattice and for several lattice sizes.}
\label{fig:best2}
\end{figure}
%%%%%%%%%%%%%%%%%%%%%%%%%%%%%%%%%%%%%%%%%%%%%%%%%%%%%%%%%%%%%%%%%%%%%%

\figurename~\ref{fig:best2} shows equivalent information, possibly in a more useful format: 
we consider again all lattices showing $T_s$, the time (in nsec) 
required to handle {\em one} lattice point by {\em one} GPU: we see 
an abrupt (and expected) transition as we move from 1 GPU to more GPUs, 
then a large plateau for large lattices and gentle scaling violations for the smaller lattices.

\begin{figure}[t]
\centering
\includegraphics[width=\textwidth]{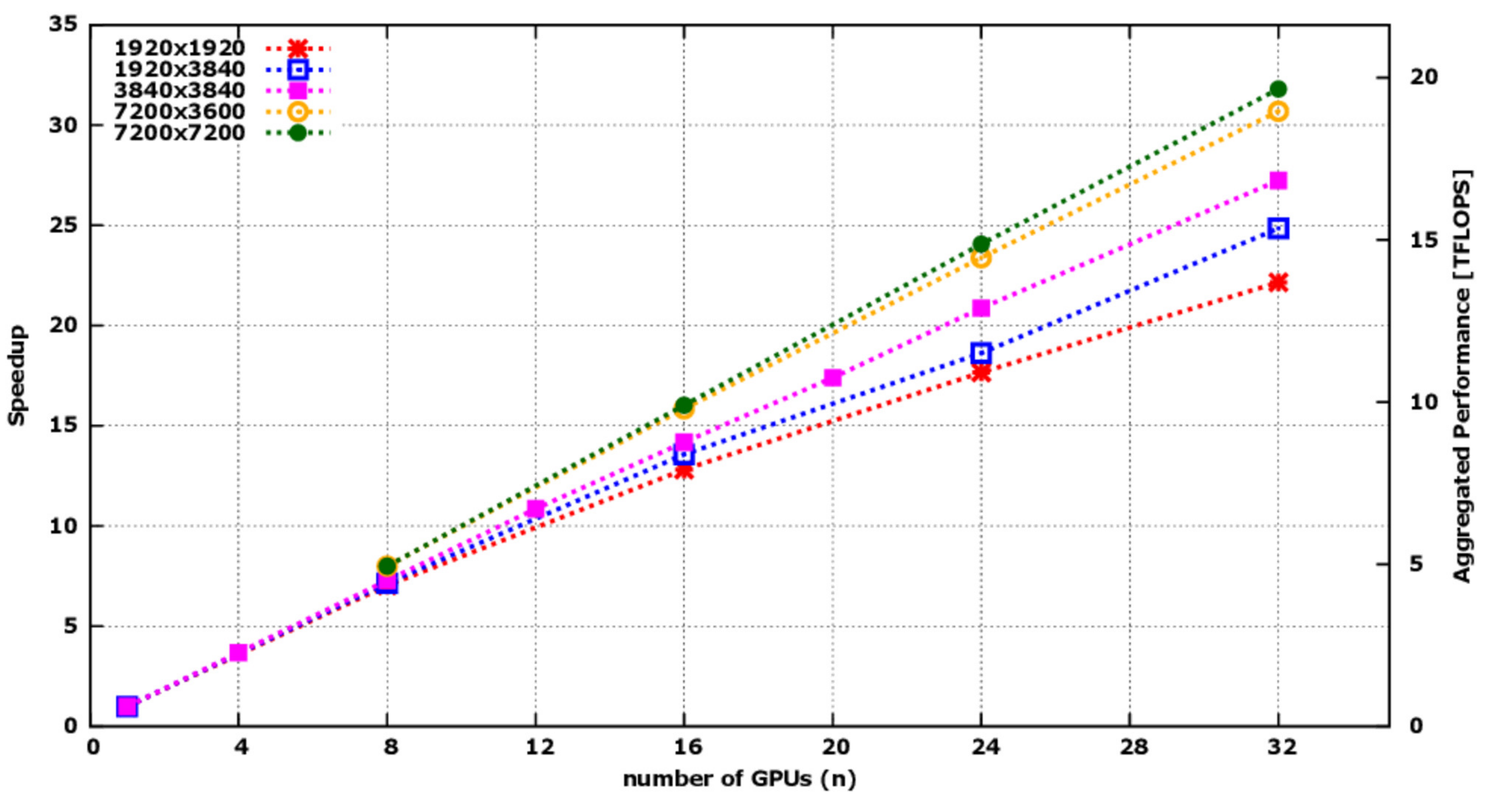}
\caption{Aggregate performance of the code for the best lattice tiling 
as a function of the number of GPUs, for several lattice sizes. Results 
are shown as speedup values (left) or effective sustained performance (Tflops, right). }
\label{fig:combo}
\end{figure}

For physics users, the ultimate metric is 
the relative speed up and the effective performance as a function 
of $n$; this is shown in \figurename~\ref{fig:combo}, that wraps up
our results. 
%
%%%%%%%%%%%%%%%%%%%%%%%%%%%%%%%%%%%%%%%%%%%%%%%%%%%%%%%%%%%%%%%%%%%%%%
%%%%%%%%%%%%%%%%%%%%%%%%%%%%%%%%%%%%%%%%%%%%%%%%%%%%%%%%%%%%%%%%%%%%%%
%
Performance increases smoothly for all lattices and number of GPUs: performance is not ideal but the bottom line of this analysis
is that our codes run efficiently on up to at least 32 GPUs for 
physics relevant lattice sizes, with a sustained performance in the range
of tens of Tflops.

\subsubsection*{Overview of Physics Results}

\begin{figure}
\includegraphics[width=0.24\textwidth]{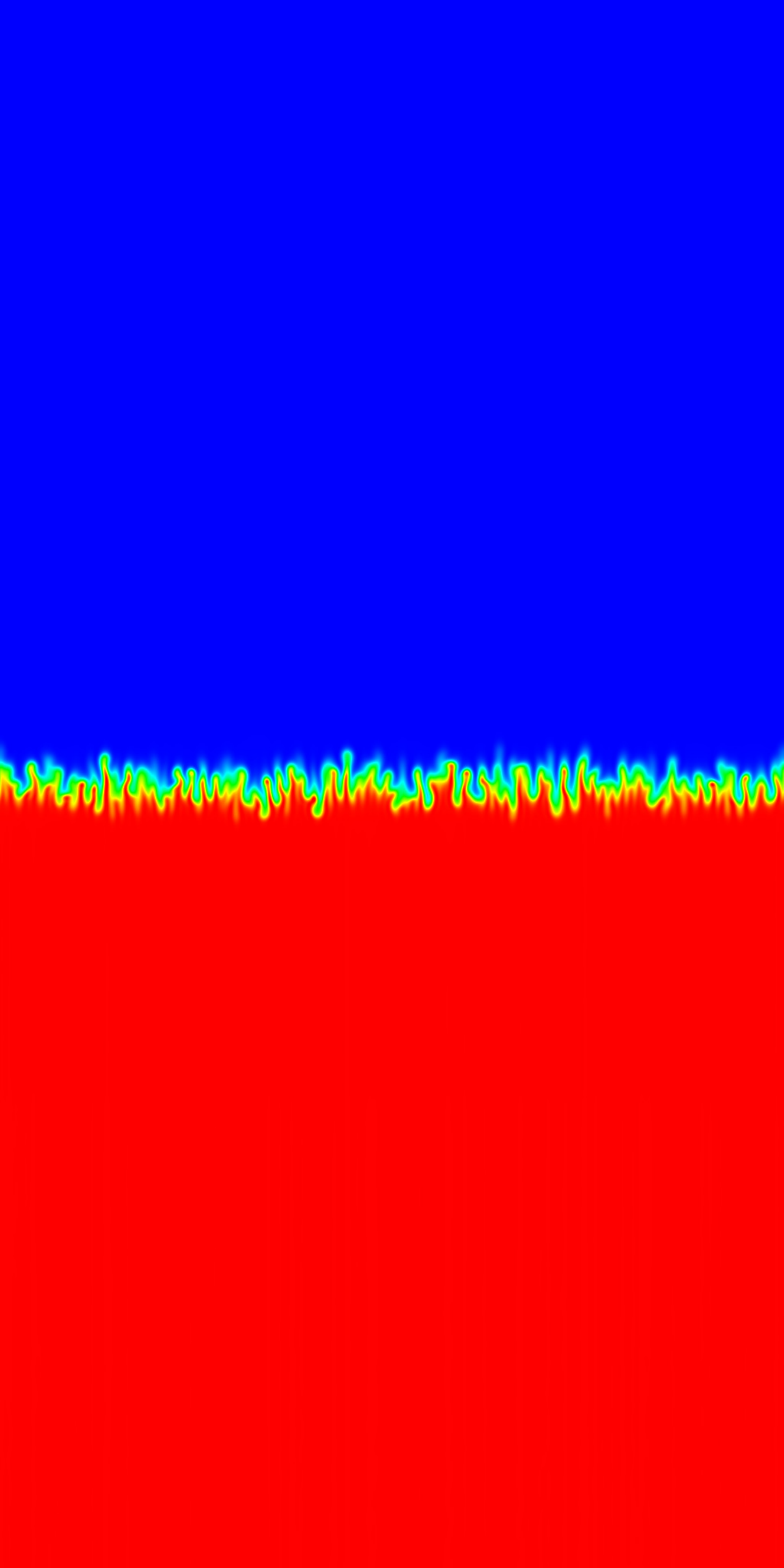}
\includegraphics[width=0.24\textwidth]{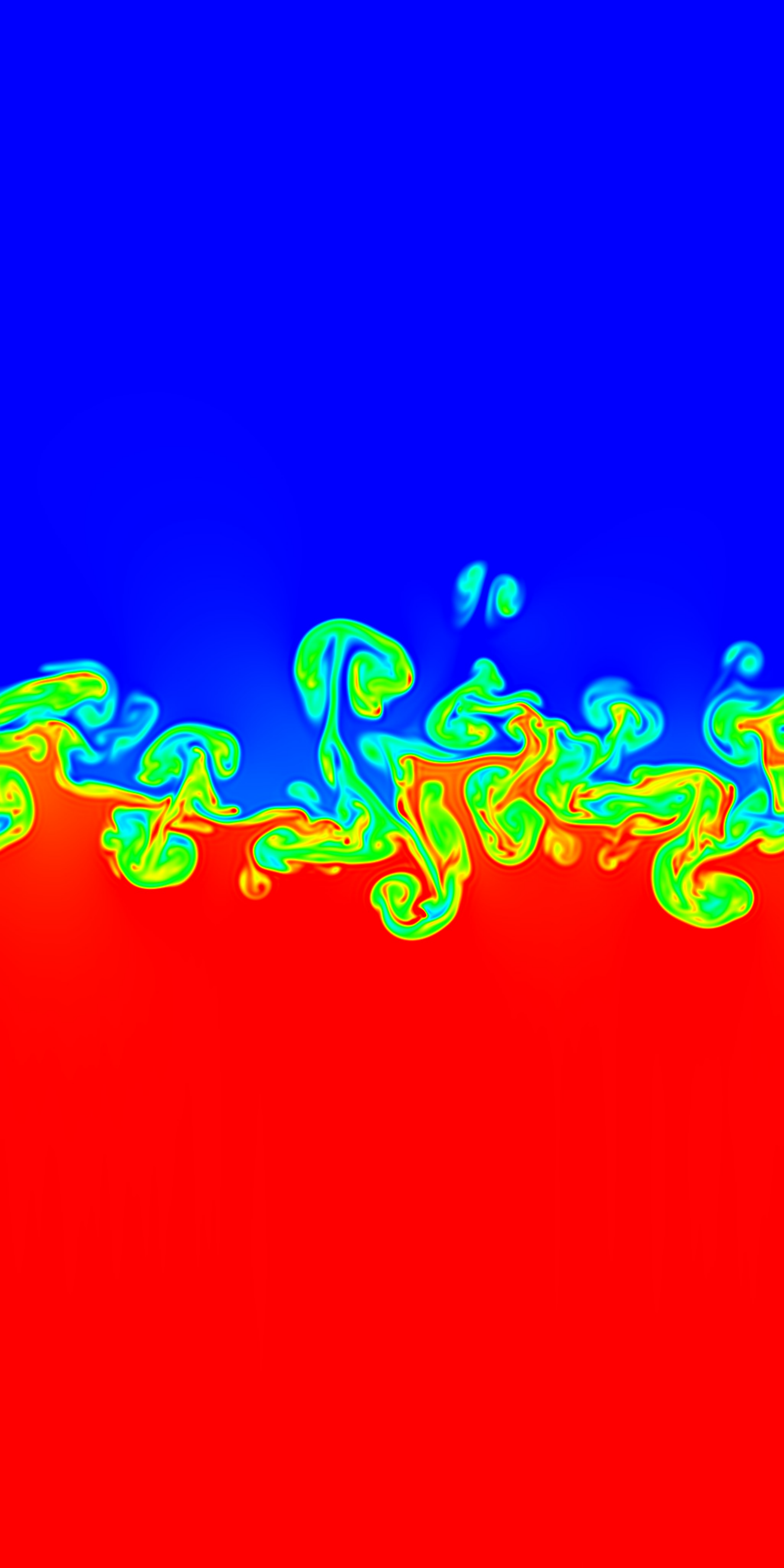}
\includegraphics[width=0.24\textwidth]{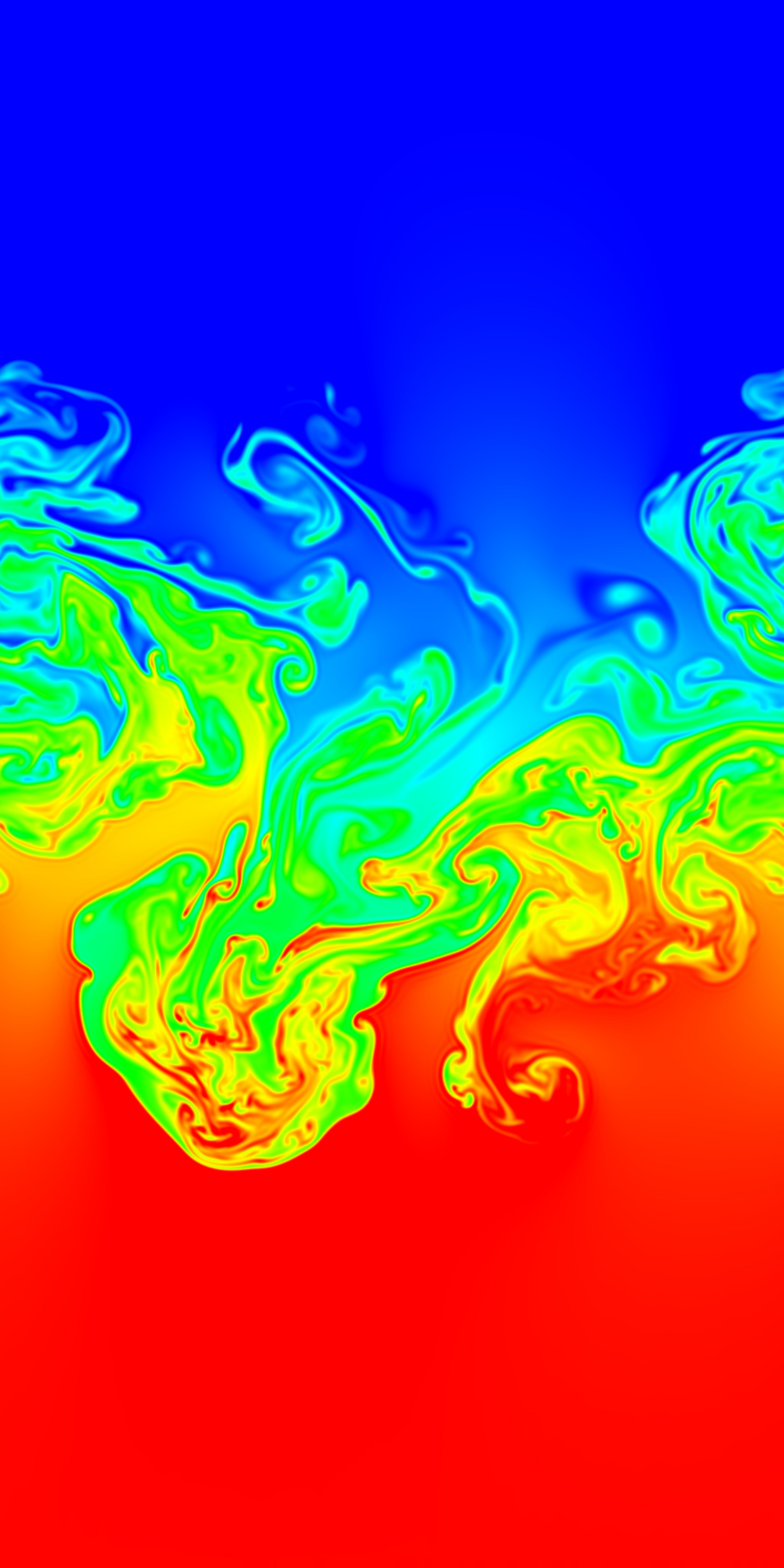}
\includegraphics[width=0.24\textwidth]{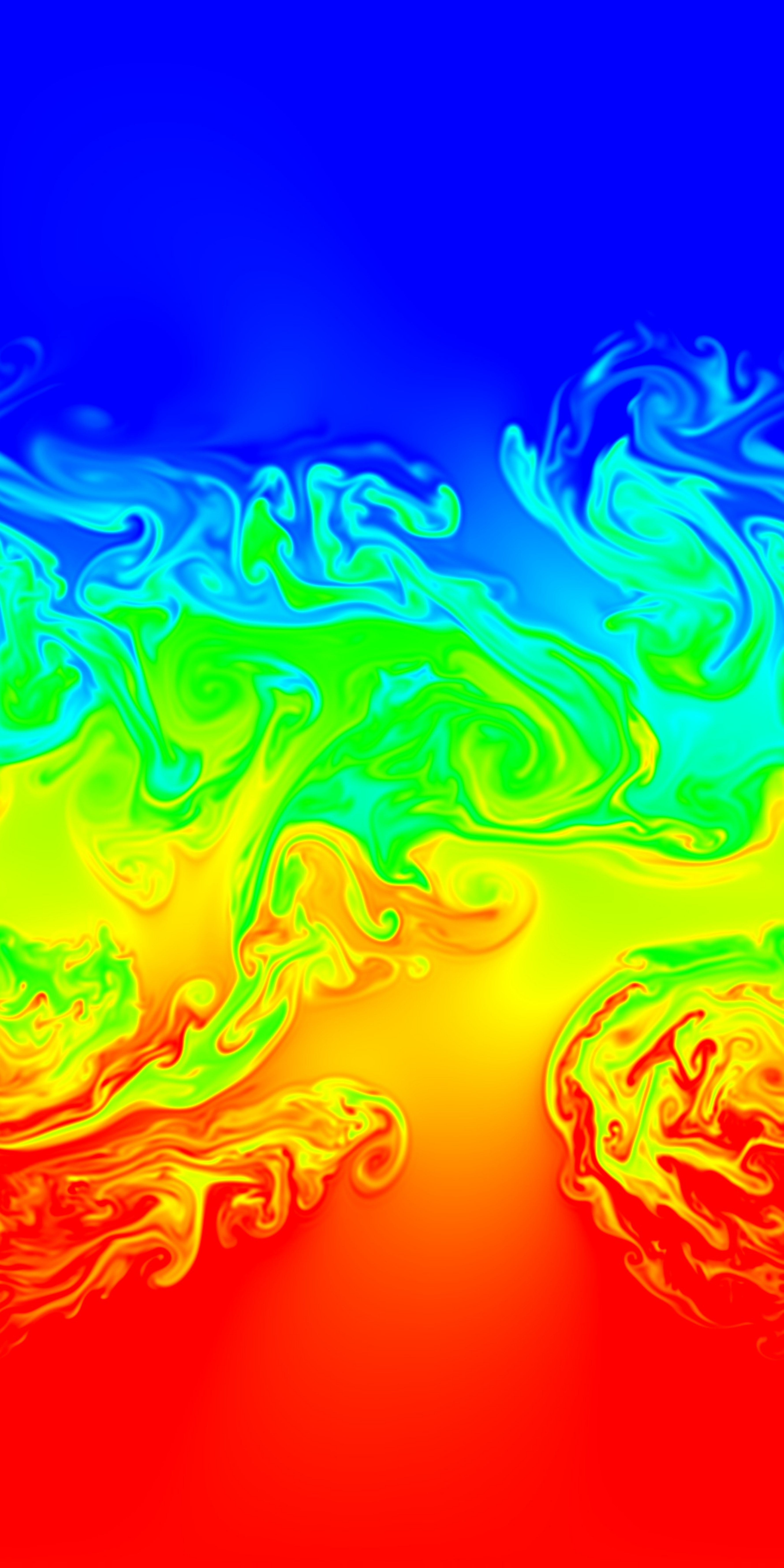}
\caption{\label{simulation}Temperature maps (red: highest temperature, 
blue: lowest temperature) of a simulation of a Rayleigh-Taylor 
instability on a lattice of $2048 \times 4096$ cells at several stages 
of its time evolution.}
\end{figure}

We finally mention that the codes described in this paper have been used 
to perform extensive studies of thermally-driven turbulence in 2D systems.
In \figurename~\ref{simulation} we show the temperature maps of a simulation of
the Rayleigh-Taylor instability at several stages  during time evolution. This
picture refers to a sample lattice of $2048 \times 4096$ cells.
Detailed physics results are in \cite{noi1, noi2, frontpropagation}.

%%%%%%%%%%%%%%%%%%%%%%%%%%%%%%%%%%%%%%%%%%%%%%%%%%%%%%%%%%%%%%%%%%%%%%%%
%%%%%%%%%%%%%%%%%%%%%%%%%%%%%%%%%%%%%%%%%%%%%%%%%%%%%%%%%%%%%%%%%%%%%%%% 

\section{Conclusions and Outlook}

This paper presents a detailed account of the development and 
optimization of a production-grade Lattice Boltzmann code on 
two recent generations of GPUs.
The strategies we have adopted are based on a quantitative approach that uses 
specific benchmarks and simple theoretical models as a guide to  
efficient implementation choices. We believe that the methodology that 
has guided our main design decisions can be helpful
to develop GPU codes for other scientific applications.

We have obtained excellent sustained performance. 
This result admittedly builds on a carefully handcrafted adjustment 
of key kernels in the code, that takes into account the architecture 
of the target processor; however the effort of writing the corresponding 
CUDA-C code remains within reasonable limits. 

Our results build on the excellent floating point performance of GPUs and 
on the capability of the memory-interface to support a large 
fraction of the peak bandwidth if one pays appropriate attention 
to the memory access patterns.

When going to a large multi-GPU implementation, node-to-node communication 
quickly become a serious bottleneck, so an optimization effort is needed 
both at the level of the algorithm and of the communication tools; 
algorithmic optimizations strictly depend on the application, while for 
the second point the CUDA-aware supports available on MPI 
libraries has significant benefits.

For problem sizes relevant for physics, fairly large systems (e.g. 32 GPUs) 
have very good strong scaling results, and the aggregated 
performance is not too far from $\approx 20$ Tflops.

Take-away lessons, applicable to different applications and beyond 
the obvious fact that as much parallelism as possible must be exposed, 
are:
\benu
\item performance depends sharply on the number of used threads; this has 
      a big impact on the computing structure of the code which, in the case of GPUs, must have a high level of data vectorization;  
\item good data allocation strategies help the memory controller 
      coalesce memory requests; for GPUs the 
      SoA scheme is preferred;
      this may require a complete or partial rewriting of existing
      applications since many codes use the AoS scheme 
      which better fits the cache structure of traditional CPUs;
\item data transfers in and out of the accelerator must be minimized and 
      overlapped with computation. Currently this is a serious 
      limit to strong scalability for applications running on GPU clusters.
      CUDA-aware MPI library implementations, enabling specific GPU supports
       (CUDA-IPC and GPUDirect RDMA) provide significant advantages. 
\eenu

A number of programming frameworks have been introduced recently, with the 
aim to help programmers to write ``portable'' codes. 
Some of these like OpenCL use a language approach like CUDA, but they 
target a wider range of accelerator architectures. 
Other frameworks, e.g., OpenMP4 and OpenACC, use directives, allowing programmers to annotate regions of codes where parallelism can be exploited, so they can be automatically mapped and optimized by a compiler for several target 
parallel architectures.
These environments are still immature in some respects: i) 
they do not reach yet the same level of performance as codes written 
using specific languages for the target accelerator, or ii) they support 
efficiently just a limited subset of architectures, leaving the portability 
issues partially solved.
We are currently investigating the advantages of these frameworks for LBM 
code developments; for preliminary results see~\cite{iccs14,europar14,europar15,jiri,cpe2016}.

%%%%%%%%%%%%%%%%%%%%%%%%%%%%%%%%%%%%%%%%%%%%%%%%%%%%%%%%%%%%%%%%%%%%%%%%

\section*{Acknowledgments}
We would like to acknowledge the support of the NVIDIA-lab at 
the J\"ulich Supercomputing Centre (J\"ulich, Germany), and of 
the NVIDIA Technology Center to allow us to use the cluster of K80 GPUs.  
This work has been done in the framework of the {\em COKA}, {\em COSA} 
and {\em SUMA} projects of INFN (Italy). We thank GF. Bilardi for useful comments;
AG has been supported by the European Union's Horizon 2020 research and
innovation programme under the Marie Sklodowska-Curie grant agreement No 642069.
%A.~G. has been supported by the 
%HPC-LEAP EJD programme of the EU (grant number 642069). 

%%%%%%%%%%%%%%%%%%%%%%%%%%%%%%%%%%%%%%%%%%%%%%%%%%%%%%%%%%%%%%%%%%%%%%%%

\bibliographystyle{elsarticle-num}
\bibliography{biblio,other,online}

\begin{thebibliography}{10}
\expandafter\ifx\csname url\endcsname\relax
  \def\url#1{\texttt{#1}}\fi
\expandafter\ifx\csname urlprefix\endcsname\relax\def\urlprefix{URL }\fi
\expandafter\ifx\csname href\endcsname\relax
  \def\href#1#2{#2} \def\path#1{#1}\fi

\bibitem{stencil1}
N.~Maruyama, T.~Aoki, Optimizing stencil computations for nvidia kepler gpus,
  in: International Workshop on High-Performance Stencil Computations, 2014,
  pp. 1--7.

\bibitem{stencil2}
A.~Vizitiu, L.~Itu, L.~Lazar, C.~Suciu, Double precision stencil computations
  on kepler gpus, in: System Theory, Control and Computing (ICSTCC), 2014 18th
  International Conference, 2014, pp. 123--127.
\newblock \href {http://dx.doi.org/10.1109/ICSTCC.2014.6982402}
  {\path{doi:10.1109/ICSTCC.2014.6982402}}.

\bibitem{stencil3}
J.~Holewinski, L.-N. Pouchet, P.~Sadayappan, High-performance code generation
  for stencil computations on gpu architectures, in: Proceedings of the 26th
  ACM International Conference on Supercomputing, ICS '12, ACM, New York, NY,
  USA, 2012, pp. 311--320.
\newblock \href {http://dx.doi.org/10.1145/2304576.2304619}
  {\path{doi:10.1145/2304576.2304619}}.

\bibitem{stencil4}
A.~Vizitiu, L.~Itu, C.~Niţă, C.~Suciu, Optimized three-dimensional stencil
  computation on fermi and kepler gpus, in: High Performance Extreme Computing
  Conference (HPEC), 2014 IEEE, 2014, pp. 1--6.
\newblock \href {http://dx.doi.org/10.1109/HPEC.2014.7040968}
  {\path{doi:10.1109/HPEC.2014.7040968}}.

\bibitem{cudagpucode}
C.~Bonati, G.~Cossu, M.~D'Elia, P.~Incardona, {QCD} simulations with staggered
  fermions on {GPUs}, Computer Physics Communications 183~(4) (2012) 853--863.
\newblock \href {http://dx.doi.org/10.1016/j.cpc.2011.12.011}
  {\path{doi:10.1016/j.cpc.2011.12.011}}.

\bibitem{QUDA}
M.~A. Clark, R.~Babich, K.~Barros, R.~C. Brower, C.~Rebbi, {Solving Lattice QCD
  systems of equations using mixed precision solvers on GPUs}, Comput. Phys.
  Commun. 181 (2010) 1517--1528.
\newblock \href {http://dx.doi.org/10.1016/j.cpc.2010.05.002}
  {\path{doi:10.1016/j.cpc.2010.05.002}}.

\bibitem{toelke10}
J.~T{\"o}lke, Implementation of a lattice boltzmann kernel using the compute
  unified device architecture developed by nvidia, Computing and Visualization
  in Science 13~(1) (2008) 29--39.
\newblock \href {http://dx.doi.org/10.1007/s00791-008-0120-2}
  {\path{doi:10.1007/s00791-008-0120-2}}.

\bibitem{bernaschi10}
M.~Bernaschi, M.~Fatica, S.~Melchionna, S.~Succi, E.~Kaxiras, A flexible
  high-performance lattice boltzmann gpu code for the simulations of fluid
  flows in complex geometries, Concurrency and Computation: Practice and
  Experience 22~(1) (2010) 1--14.
\newblock \href {http://dx.doi.org/10.1002/cpe.1466}
  {\path{doi:10.1002/cpe.1466}}.

\bibitem{ripesi14}
P.~Ripesi, L.~Biferale, S.~F. Schifano, R.~Tripiccione, Evolution of a
  double-front rayleigh-taylor system using a graphics-processing-unit-based
  high-resolution thermal lattice-boltzmann model, Physical Review E -
  Statistical, Nonlinear, and Soft Matter Physics 89~(4).
\newblock \href {http://dx.doi.org/10.1103/PhysRevE.89.043022}
  {\path{doi:10.1103/PhysRevE.89.043022}}.

\bibitem{iccs10}
L.~Biferale, F.~Mantovani, M.~Pivanti, M.~Sbragaglia, A.~Scagliarini, S.~F.
  Schifano, F.~Toschi, R.~Tripiccione, {Lattice Boltzmann fluid-dynamics on the
  QPACE supercomputer}, Procedia Computer Science 1~(1) (2010) 1075--1082,
  {ICCS 2010}.
\newblock \href {http://dx.doi.org/10.1016/j.procs.2010.04.119}
  {\path{doi:10.1016/j.procs.2010.04.119}}.

\bibitem{ppam11}
L.~Biferale, F.~Mantovani, M.~Pivanti, F.~Pozzati, M.~Sbragaglia,
  A.~Scagliarini, S.~F. Schifano, F.~Toschi, R.~Tripiccione, A multi-gpu
  implementation of a d2q37 lattice boltzmann code, in: R.~Wyrzykowski,
  J.~Dongarra, K.~Karczewski, J.~Wa{\'{s}}niewski (Eds.), Parallel Processing
  and Applied Mathematics: 9th International Conference, PPAM 2011, Torun,
  Poland, September 11-14, 2011. Revised Selected Papers, Part I, {Lecture
  Notes in Computer Science}, Springer Berlin Heidelberg, Berlin, Heidelberg,
  2012, pp. 640--650.
\newblock \href {http://dx.doi.org/10.1007/978-3-642-31464-3\_65}
  {\path{doi:10.1007/978-3-642-31464-3\_65}}.

\bibitem{single-proc}
G.~Wellein, T.~Zeiser, G.~Hager, S.~Donath, {On the single processor
  performance of simple lattice Boltzmann kernels}, Computers \& Fluids
  35~(8--9) (2006) 910--919, proceedings of the First International Conference
  for Mesoscopic Methods in Engineering and Science.
\newblock \href {http://dx.doi.org/10.1016/j.compfluid.2005.02.008}
  {\path{doi:10.1016/j.compfluid.2005.02.008}}.

\bibitem{Pohl}
T.~Pohl, M.~Kowarschik, J.~Wilke, K.~Iglberger, U.~R\"ude, Optimization and
  profiling of the cache performance of parallel lattice boltzmann codes,
  Parallel Processing Letters 13~(04) (2003) 549--560.
\newblock \href {http://dx.doi.org/10.1142/S0129626403001501}
  {\path{doi:10.1142/S0129626403001501}}.

\bibitem{Wittmann}
M.~Wittmann, T.~Zeiser, G.~Hager, G.~Wellein, Comparison of different
  propagation steps for the lattice boltzmann method, CoRR abs/1111.0922.

\bibitem{shet1}
A.~G. Shet, S.~H. Sorathiya, S.~Krithivasan, A.~M. Deshpande, B.~Kaul, S.~D.
  Sherlekar, S.~Ansumali, Data structure and movement for lattice-based
  simulations, Phys. Rev. E 88 (2013) 013314.
\newblock \href {http://dx.doi.org/10.1103/PhysRevE.88.013314}
  {\path{doi:10.1103/PhysRevE.88.013314}}.

\bibitem{shet2}
A.~G. Shet, K.~Siddharth, S.~H. Sorathiya, A.~M. Deshpande, S.~D. Sherlekar,
  B.~Kaul, S.~Ansumali, On vectorization for lattice based simulations,
  International Journal of Modern Physics C 24.
\newblock \href {http://dx.doi.org/10.1142/S0129183113400111}
  {\path{doi:10.1142/S0129183113400111}}.

\bibitem{sbac-pad13}
J.~Kraus, M.~Pivanti, S.~F. Schifano, R.~Tripiccione, M.~Zanella, {Benchmarking
  GPUs with a parallel Lattice-Boltzmann code}, in: {Computer Architecture and
  High Performance Computing (SBAC-PAD), 25th International Symposium on},
  IEEE, 2013, pp. 160--167.
\newblock \href {http://dx.doi.org/10.1109/SBAC-PAD.2013.37}
  {\path{doi:10.1109/SBAC-PAD.2013.37}}.

\bibitem{iccs11}
L.~Biferale, F.~Mantovani, M.~Pivanti, F.~Pozzati, M.~Sbragaglia,
  A.~Scagliarini, S.~F. Schifano, F.~Toschi, R.~Tripiccione, {Optimization of
  Multi-Phase Compressible Lattice Boltzmann Codes on Massively Parallel
  Multi-Core Systems}, Procedia Computer Science 4 (2011) 994--1003,
  proceedings of the International Conference on Computational Science, ICCS
  2011.
\newblock \href {http://dx.doi.org/10.1016/j.procs.2011.04.105}
  {\path{doi:10.1016/j.procs.2011.04.105}}.

\bibitem{caf11}
L.~Biferale, F.~Mantovani, M.~Pivanti, F.~Pozzati, M.~Sbragaglia,
  A.~Scagliarini, S.~F. Schifano, F.~Toschi, R.~Tripiccione, An optimized d2q37
  lattice boltzmann code on gp-gpus, {Computers \& Fluids} 80 (2013) 55 -- 62.
\newblock \href {http://dx.doi.org/10.1016/j.compfluid.2012.06.003}
  {\path{doi:10.1016/j.compfluid.2012.06.003}}.

\bibitem{ppam13}
M.~Pivanti, F.~Mantovani, S.~F. Schifano, R.~Tripiccione, L.~Zenesini, An
  optimized lattice boltzmann code for {BlueGene/Q}, in: R.~Wyrzykowski,
  J.~Dongarra, K.~Karczewski, J.~Wa{\'{s}}niewski (Eds.), Parallel Processing
  and Applied Mathematics: 10th International Conference, PPAM 2013, Warsaw,
  Poland, September 8-11, 2013, Revised Selected Papers, Part II, {Lecture
  Notes in Computer Science}, Springer Berlin Heidelberg, Berlin, Heidelberg,
  2014, pp. 385--394.
\newblock \href {http://dx.doi.org/10.1007/978-3-642-55195-6\_36}
  {\path{doi:10.1007/978-3-642-55195-6\_36}}.

\bibitem{sauro}
S.~Succi, {The Lattice-Boltzmann Equation}, Oxford university press, Oxford,
  2001.

\bibitem{JFM}
M.~Sbragaglia, R.~Benzi, L.~Biferale, H.~Chen, X.~Shan, S.~Succi, {Lattice
  Boltzmann method with self-consistent thermo-hydrodynamic equilibria},
  Journal of Fluid Mechanics 628 (2009) 299--309.
\newblock \href {http://dx.doi.org/10.1017/S002211200900665X}
  {\path{doi:10.1017/S002211200900665X}}.

\bibitem{POF}
A.~Scagliarini, L.~Biferale, M.~Sbragaglia, K.~Sugiyama, F.~Toschi, {Lattice
  Boltzmann methods for thermal flows: Continuum limit and applications to
  compressible Rayleigh--Taylor systems}, Physics of Fluids (1994-present)
  22~(5) (2010) 055101.
\newblock \href {http://dx.doi.org/10.1063/1.3392774}
  {\path{doi:10.1063/1.3392774}}.

\bibitem{fermi}
\href{http://www.nvidia.com/content/PDF/fermi_white_papers/NVIDIA_Fermi_Compute_Architecture_Whitepaper.pdf}{Nvidia,
  fermi}.
\newline\urlprefix\url{http://www.nvidia.com/content/PDF/fermi_white_papers/NVIDIA_Fermi_Compute_Architecture_Whitepaper.pdf}

\bibitem{kepler}
\href{http://www.nvidia.com/content/PDF/kepler/NVIDIA-Kepler-GK110-Architecture-Whitepaper.pdf}{Nvidia,
  kepler gk110}.
\newline\urlprefix\url{http://www.nvidia.com/content/PDF/kepler/NVIDIA-Kepler-GK110-Architecture-Whitepaper.pdf}

\bibitem{cuda}
\href{http://docs.nvidia.com/cuda/cuda-c-programming-guide}{Nvidia {\em cuda c
  programming guide}}.
\newline\urlprefix\url{http://docs.nvidia.com/cuda/cuda-c-programming-guide}

\bibitem{brent}
R.~P. Brent, The parallel evaluation of general arithmetic expressions, J. ACM
  21~(2) (1974) 201--206.
\newblock \href {http://dx.doi.org/10.1145/321812.321815}
  {\path{doi:10.1145/321812.321815}}.

\bibitem{cuda-aware_mpi}
\href{http://developer.nvidia.com/content/introduction-cuda-aware-mpi}{An
  introduction to cuda-aware mpi}.
\newline\urlprefix\url{http://developer.nvidia.com/content/introduction-cuda-aware-mpi}

\bibitem{rdma}
\href{http://devblogs.nvidia.com/parallelforall/benchmarking-gpudirect-rdma-on-modern-server-platforms}{Benchmarking
  gpudirect rdma on modern server platforms}.
\newline\urlprefix\url{http://devblogs.nvidia.com/parallelforall/benchmarking-gpudirect-rdma-on-modern-server-platforms}

\bibitem{caf13}
F.~Mantovani, M.~Pivanti, S.~F. Schifano, R.~Tripiccione, Performance issues on
  many-core processors: A d2q37 lattice boltzmann scheme as a test-case,
  {Computers \& Fluids} 88 (2013) 743 -- 752.
\newblock \href {http://dx.doi.org/10.1016/j.compfluid.2013.05.014}
  {\path{doi:10.1016/j.compfluid.2013.05.014}}.

\bibitem{ccp12}
F.~Mantovani, M.~Pivanti, S.~F. Schifano, R.~Tripiccione, Exploiting
  parallelism in many-core architectures: Lattice boltzmann models as a test
  case, Journal of Physics: Conference Series 454~(1).
\newblock \href {http://dx.doi.org/10.1088/1742-6596/454/1/012015}
  {\path{doi:10.1088/1742-6596/454/1/012015}}.

\bibitem{iccs13}
G.~Crimi, F.~Mantovani, M.~Pivanti, S.~F. Schifano, R.~Tripiccione, {Early
  Experience on Porting and Running a Lattice Boltzmann Code on the Xeon-phi
  Co-Processor}, Procedia Computer Science 18 (2013) 551--560.
\newblock \href {http://dx.doi.org/10.1016/j.procs.2013.05.219}
  {\path{doi:10.1016/j.procs.2013.05.219}}.

\bibitem{chep13}
G.~Bortolotti, M.~Caberletti, G.~Crimi, A.~Ferraro, F.~Giacomini, M.~Manzali,
  G.~Maron, M.~Pivanti, D.~Salomoni, S.~F. Schifano, R.~Tripiccione,
  M.~Zanella, Computing on knights and kepler architectures, Journal of
  Physics: Conference Series 513~(5) (2014) 052032.

\bibitem{noi1}
L.~Biferale, F.~Mantovani, M.~Sbragaglia, A.~Scagliarini, F.~Toschi,
  R.~Tripiccione, {Second-order closure in stratified turbulence: Simulations
  and modeling of bulk and entrainment regions}, Physical Review E 84~(1)
  (2011) 016305.
\newblock \href {http://dx.doi.org/10.1103/PhysRevE.84.016305}
  {\path{doi:10.1103/PhysRevE.84.016305}}.

\bibitem{noi2}
L.~Biferale, F.~Mantovani, M.~Sbragaglia, A.~Scagliarini, F.~Toschi,
  R.~Tripiccione, {Reactive Rayleigh-Taylor systems: Front propagation and
  non-stationarity}, EPL 94~(5) (2011) 54004.
\newblock \href {http://dx.doi.org/10.1209/0295-5075/94/54004}
  {\path{doi:10.1209/0295-5075/94/54004}}.

\bibitem{frontpropagation}
A.~Scagliarini, L.~Biferale, F.~Mantovani, M.~Pivanti, F.~Pozzati,
  M.~Sbragaglia, S.~F. Schifano, F.~Toschi, R.~Tripiccione, {Front propagation
  in Rayleigh-Taylor systems with reaction}, in: {Journal of Physics:
  Conference Series}, Vol. 318, IOP Publishing, 2011, pp. 1--10.
\newblock \href {http://dx.doi.org/10.1088/1742-6596/318/9/092024}
  {\path{doi:10.1088/1742-6596/318/9/092024}}.

\bibitem{iccs14}
E.~Calore, S.~F. Schifano, R.~Tripiccione, {A Portable OpenCL Lattice Boltzmann
  Code for Multi-and Many-core Processor Architectures}, Procedia Computer
  Science 29 (2014) 40--49.
\newblock \href {http://dx.doi.org/10.1016/j.procs.2014.05.004}
  {\path{doi:10.1016/j.procs.2014.05.004}}.

\bibitem{europar14}
E.~Calore, S.~F. Schifano, R.~Tripiccione, On portability, performance and
  scalability of an mpi opencl lattice boltzmann code, in: L.~Lopes,
  J.~{\v{Z}}ilinskas, A.~Costan, R.~G. Cascella, G.~Kecskemeti, E.~Jeannot,
  M.~Cannataro, L.~Ricci, S.~Benkner, S.~Petit, V.~Scarano, J.~Gracia,
  S.~Hunold, S.~L. Scott, S.~Lankes, C.~Lengauer, J.~Carretero, J.~Breitbart,
  M.~Alexander (Eds.), Euro-Par 2014: Parallel Processing Workshops: Euro-Par
  2014 International Workshops, Porto, Portugal, August 25-26, 2014, Revised
  Selected Papers, Part II, {Lecture Notes in Computer Science}, Springer
  International Publishing, Cham, 2014, pp. 438--449.
\newblock \href {http://dx.doi.org/10.1007/978-3-319-14313-2\_37}
  {\path{doi:10.1007/978-3-319-14313-2\_37}}.

\bibitem{europar15}
E.~Calore, J.~Kraus, S.~F. Schifano, R.~Tripiccione, Accelerating lattice
  boltzmann applications with openacc, in: L.~J. Tr{\"a}ff, S.~Hunold,
  F.~Versaci (Eds.), Euro-Par 2015: Parallel Processing: 21st International
  Conference on Parallel and Distributed Computing, Vienna, Austria, August
  24-28, 2015, Proceedings, {Lecture Notes in Computer Science}, Springer
  Berlin Heidelberg, Berlin, Heidelberg, 2015, pp. 613--624.
\newblock \href {http://dx.doi.org/10.1007/978-3-662-48096-0\_47}
  {\path{doi:10.1007/978-3-662-48096-0\_47}}.

\bibitem{jiri}
J.~Kraus, M.~Schlottke, A.~Adinetz, D.~Pleiter, Accelerating a c++ cfd code
  with openacc, in: Accelerator Programming using Directives (WACCPD), 2014
  First Workshop on, 2014, pp. 47--54.
\newblock \href {http://dx.doi.org/10.1109/WACCPD.2014.11}
  {\path{doi:10.1109/WACCPD.2014.11}}.

\bibitem{cpe2016}
E.~Calore, A.~Gabbana, J.~Kraus, S.~F. Schifano, R.~Tripiccione, Performance
  and portability of accelerated lattice boltzmann applications with openacc,
  Concurrency and Computation: Practice and Experience\href
  {http://dx.doi.org/10.1002/cpe.3862} {\path{doi:10.1002/cpe.3862}}.

\end{thebibliography}

\newpage

%%%%%%%%%%%%%%%%%%%%%%%%%%%%%%%%%%%%%%%%%%%%%%%%%%%%%%%%%%%%%%%%%%%%%%%%

\appendix\section{Code Listings}

%%%%%%%%%%%%%%%%%%%%%%%%%%%%%%%%%%%%%%%%%%%%%%%%%%%%%%%%%%%%%%%%%%%%%%%%
%
\begin{figure}[h]
\centering
\begin{lstlisting}[basicstyle=\scriptsize]
// Pack non-contiguous buffer in a contiguous buffer
__global__ void pack(data_t *f, data_t *sndBuf){
  int idx_c, idx_nc;  
  idx_c  = ( blockIdx.y  * blockDim.y * blockDim.x ) +  
           ( threadIdx.y * blockDim.x              ) +  
           ( threadIdx.x                           );
  idx_nc = ( blockIdx.y  * blockDim.y * NY ) +  
           ( threadIdx.y * NY              ) +  
           ( threadIdx.x                   );    
  if( threadIdx.x < blockDim.x )
    sndBuf[ idx_c ] = f[ idx_nc ];
}

// Unpack a contiguous buffer in a non-contiguous buffer
__global__ void unpack(data_t *rcvBuf, data_t *f){
  int idx_c, idx_nc;
  idx_c  = ( blockIdx.y  * blockDim.y * blockDim.x ) +
           ( threadIdx.y * blockDim.x              ) +
           ( threadIdx.x                           ); 
  idx_nc = ( blockIdx.y  * blockDim.y * NY ) + 
           ( threadIdx.y * NY              ) + 
           ( threadIdx.x                   );
  if( threadIdx.x < blockDim.x )
    f[ idx_nc ] = rcvBuf[ idx_c ];
}
\end{lstlisting}
\caption{\label{fig:pack-unpack}
CUDA implementation of the {\tt pack} and {\tt unpack} kernels. 
{\tt pack} starts a CUDA grid, 
and each thread reads a data item from a lattice site and stores it into 
an array at consecutive address. {\tt unpack} does the opposite.
}
\end{figure}

%%%%%%%%%%%%%%%%%%%%%%%%%%%%%%%%%%%%%%%%%%%%%%%%%%%%%%%%%%%%%%%%%%%%%%%%

\newpage

%%%%%%%%%%%%%%%%%%%%%%%%%%%%%%%%%%%%%%%%%%%%%%%%%%%%%%%%%%%%%%%%%%%%%%%%
%
\begin{figure}
\centering
\begin{lstlisting}[basicstyle=\scriptsize]
// send/receive buffer allocation
cudaMalloc((void **) &sndTopBuf, LY*HX*sizeof(data_t));
cudaMalloc((void **) &rcvBotBuf, LY*HX*sizeof(data_t)); 
cudaMalloc((void **) &sndBotBuf, LY*HX*sizeof(data_t));
cudaMalloc((void **) &rcvTopBuf, LY*HX*sizeof(data_t)); 

// function to transfer non-contiguous borders
void pbc_nc() {

  // Pack top non-contiguous border
  pack <<<dimGrid, dimBlock>>> (src_p, sndTopBuf); 

  // Wait the end of pack-kernel
  cudaDeviceSynchronize();                      

  // Exchange the contiguous buffer using CUDA-aware MPI
  MPI_Sendrecv(
    sndTopBuf, LY*HX, MPI_DOUBLE, mpi_left, 0,
    rcvBotBuf, LY*HX, MPI_DOUBLE, mpi_right, 0,
    MPI_COMM_WORLD, MPI_STATUS_IGNORE
  );

  // Unpack data on bottom halo
  unpack <<< dimGrid, dimBlock >>> (rcvBotBuf, dst_p); 
  
  
  // Pack bottom non-contiguous border
  pack <<<dimGrid, dimBlock>>> (src_p, sndBotBuf); 

  // Wait the end of pack-kernel
  cudaDeviceSynchronize();                      

  // Exchange the contiguous buffer using CUDA-aware MPI
  MPI_Sendrecv(
    sndBotBuf, LY*HX, MPI_DOUBLE, mpi_left, 0,
    rcvTopBuf, LY*HX, MPI_DOUBLE, mpi_right, 0,
    MPI_COMM_WORLD, MPI_STATUS_IGNORE
  );

  // Unpack data on bottom halo
  unpack <<< dimGrid, dimBlock >>> (rcvTopBuf, dst_p); 
  
  // Wait end of unpack kernel
  cudaDeviceSynchronize();                          
}
\end{lstlisting}
\caption{\label{fig:nc-buffers}
Sample code to handle 
data transfers among non contiguous buffers. Two buffers {\tt sndBuf} and {\tt rcvBuf} 
are persistently allocated on the GPU. Function {\tt pbc\_nc} performs the 
transfers of the data associated to the left and right halos.
Real code we use is more complex because we overlaps the two transfers as 
much as possible running pack and unpack kernels on separate streams, and 
using asynchronous MPI operations. 
Execution of this function has to be completed before starting operations 
on lattice bulk and update of contiguous halos to ensure that halos are 
correctly updated.
}
\end{figure}

%%%%%%%%%%%%%%%%%%%%%%%%%%%%%%%%%%%%%%%%%%%%%%%%%%%%%%%%%%%%%%%%%%%%%%%%

\newpage

%%%%%%%%%%%%%%%%%%%%%%%%%%%%%%%%%%%%%%%%%%%%%%%%%%%%%%%%%%%%%%%%%%%%%%%%
%
\begin{figure}
\centering
\begin{lstlisting}[basicstyle=\tiny]
// update non-contiguous halos
pbc_nc( f2_soa_d );

// pack right/left borders
pack_right <<< ..., stream[1] >>> ( ... );    
pack_left  <<< ..., stream[2] >>> ( ... ); 

// run propagateAndCollide over Bulk
propagateCollideBulk <<< ..., stream[0] >>> ( ... );

// wait end of pack right borders
cudaStreamSynchronize(stream[1]);
// perform MPI operations
MPI_Sendrecv( );     

// wait end of pack left borders
cudaStreamSynchronize(stream[2]);
MPI_Sendrecv( );

// unpack right/left halos
unpack_left  <<< ..., stream[1] >>> ( ... );    
unpack_right <<< ..., stream[2] >>> ( ... );

// wait end of unpack right/left halos 
// (required before starting processing left, right, top and bottom borders)
cudaStreamSynchronize(stream[1]);
cudaStreamSynchronize(stream[2]);

// process left/right borders
propagateCollideL <<< ..., stream[1] >>> ( ... );    
propagateCollideR <<< ..., stream[2] >>> ( ... );

// process top/bottom borders
if (uppermost-rank){
  propagateT <<< ..., stream[3] >>> ( ... );    
  bcT        <<< ..., stream[3] >>> ( ... );    
  collideT   <<< ..., stream[3] >>> ( ... );
} else { 
  propagateCollideT <<< ..., stream[3] >>> ( ... );
}

if (lowermost-rank){
  propagateB <<< ..., stream[4] >>> ( ... );    
  bcB        <<< ..., stream[4] >>> ( ... );    
  collideB   <<< ..., stream[4] >>> ( ... );
} else { 
  propagateCollideB <<< ..., stream[4] >>> ( ... );
}

cudaDeviceSynchronize();
\end{lstlisting}
\caption{\label{fig:2D-code-with-overlap-fused}
Overall organization of the code for a 2-D tiling of the lattice, fusing
the {\tt propagate} and {\tt collide} kernels in one step.
The code first update non contiguous halos calling function {\tt pbc\_nc} 
explained in \figurename~\ref{fig:nc-buffers}.
After this is fully completed, starts pack of left and right borders on two 
GPU streams, and in parallel starts execution of {\tt propagateCollideBulk} 
kernel processing lattice bulk.
As MPI operations complete, data are unpacked on left and right halos, and 
we start processing of left, right, top and bottom borders.
For left and right borders, we apply kernels {\tt propagateCollideL} and 
{\tt propagateCollideR} computing propagate and collide phases in one step.
For top and bottom borders, we apply in sequence {\tt propgate}, {\tt bc} and 
{\tt collide} kernels is the GPU is associated to a tile at top and bottom 
region of the lattice. Otherwise we only run {\tt propagateCollideT} and 
{\tt propagateCollideB}.
}
\end{figure}

%%%%%%%%%%%%%%%%%%%%%%%%%%%%%%%%%%%%%%%%%%%%%%%%%%%%%%%%%%%%%%%%%%%%%%%%

\end{document}